\documentclass[lettersize,journal]{IEEEtran}
\usepackage{amsmath,amsfonts}
\usepackage{algorithmic}
\usepackage{algorithm}
\usepackage{array}

\usepackage[caption=false,font=footnotesize,labelfont=rm,textfont=rm]{subfig}

\usepackage{textcomp}
\usepackage{stfloats}
\usepackage{url}
\usepackage{verbatim}
\usepackage{graphicx}
\usepackage{bm}
\usepackage{cite}
\usepackage{subcaption}

\usepackage{xcolor}

\begin{document}

\title{NF-SecRIS: RIS-Assisted Near-Field Physical Layer Security via Secure Location Modulation}

\author{Zhendong Wang, Chenyang Meng, Jun Yang, Jiayuan Wang, Yin Li, Linshan Jiang, Jin Zhang
\thanks{Manuscript received 01 November 2025. (\textit{Corresponding author: Jin Zhang.})

Zhendong Wang is with the Research Institute of Trustworthy Autonomous Systems, Department of Computer Science and Engineering, Southern University of Science and Technology, Shenzhen 518055, China and also with Department of Broadband Communication, Peng Cheng Laboratory, Shenzhen, China (e-mail: 12131106@mail.sustech.edu.cn).

Chenyang Meng, Jun Yang, and Yin Li are with Department of Broadband Communication, Peng Cheng Laboratory, Shenzhen, China.

Linshan Jiang is with Institute of Data Science at National University of Singapore (e-mail: linshan@nus.edu.sg).

Jiayuan Wang and Jin Zhang are with the Research Institute of Trustworthy Autonomous Systems, Department of Computer Science and Engineering, Southern University of Science and Technology, Shenzhen 518055, China (e-mail: 12540046@mail.sustech.edu.cn; zhangj4@sustech.edu.cn)}}

\markboth{}%
{Shell \MakeLowercase{\textit{et al.}}: A Sample Article Using IEEEtran.cls for IEEE Journals}


\maketitle

\begin{abstract}
The 6G wireless networks impose extremely high requirements on physical layer secure communication. However, the existing solutions usually can only achieve one-dimensional physical layer security (PLS) in the angle dimension, and cannot achieve PLS in the range dimension. In this paper, we propose the NF-SecRIS system, the first range-angle-dependent (2D) PLS near-field communication system based on ultra-large-scale reconfigurable intelligent surface (RIS). We propose the secure location modulation scheme to synthesize the near-field spatial-temporal coding pattern of RIS with extremely low complexity. It ensures that only legitimate user can receive the raw constellations, while potential eavesdroppers at other ranges or angles can only receive the obfuscated constellations. NF-SecRIS operates without requiring synchronization with either transmitter or receiver. We implement a prototype of NF-SecRIS and conduct comprehensive experiments with multiple modulation schemes. The results show that the bit error rate (BER) of legitimate user is below $10^{-4}$, while eavesdroppers at other ranges or angles suffer from BER exceeding $40\%$. It validates the implementation of 2D PLS in near-field communications.


\end{abstract}


\section{Introduction}

\IEEEPARstart{I}{n} the upcoming 6G era, wireless networks are expected to exhibit multi-functional capabilities, among which integrating communication and physical layer security (PLS) has drawn significant attentions \cite{9755276, 10720877,11103477,10718344}. PLS techniques leverage the inherent properties of communication channels to ensure that legitimate user and eavesdroppers receive distinctly different signals, thereby achieving secure communications \cite{1055917,7762075}. Compared to cryptographic encryption scheme, PLS has the fundamental advantage of preventing eavesdroppers from leveraging powerful computational resources to conduct brute-force cracking.

Most existing PLS systems support only secure communications in the angular domain and fail to prevent eavesdropping along the range dimension \cite{10.1145/3495243.3560547,s41467-025-60725-1,s41467-025-63326-0,10.1038/s41467-024-50482-y,10.1038/s41928-021-00664-z,10.1063/5.0132854,8334230}. However, the ideal situation of PLS is that only the legitimate user can receive the raw constellations, while eavesdroppers at other ranges or angles can only receive the obfuscated constellations resulting no capability to recover the raw data, that is known as range-angle-dependent (2D) secure communications \cite{8486339}. Although the study in \cite{8486339} has achieved 2D secure communications, its distributed far-field architecture requires strict synchronization among multiple transmitters and introduces high complexity and high cost. To avoid the limitations of distributed architecture,
the single transmitter must have the capability to simultaneously manipulate electromagnetic wave in both range domain and angle domain \cite{1143539}. Nevertheless, the traditional far-field techniques lack the manipulation ability in range domain. To address this, the unique properties of near-field have recently attracted considerable interests due to the range-angle dependency of spherical wavefront \cite{10944643,10436390,10684477,10934753,10971913}. To guarantee that the user operates within the near-field region of transmitter, an ultra-large-scale array is required, because the near-field range is positively correlated with the array aperture \cite{10716601,10.5555/3388242.3388317,10541333}. With the development of electromagnetic array technology, the reconfigurable intelligent surface (RIS) outperforms conventional phased array in ultra-large-scale deployment owing to key advantages such as simple structure, low cost, wireless feeding capability, and low power consumption \cite{9424177,9140329}. Hence, integrating near-field communication and 2D PLS with the assistance of ultra-large-scale RIS is a promising research area with significant potential \cite{10833623,10380596}.

\begin{figure}[!t]
\centering
\includegraphics[width=250pt]{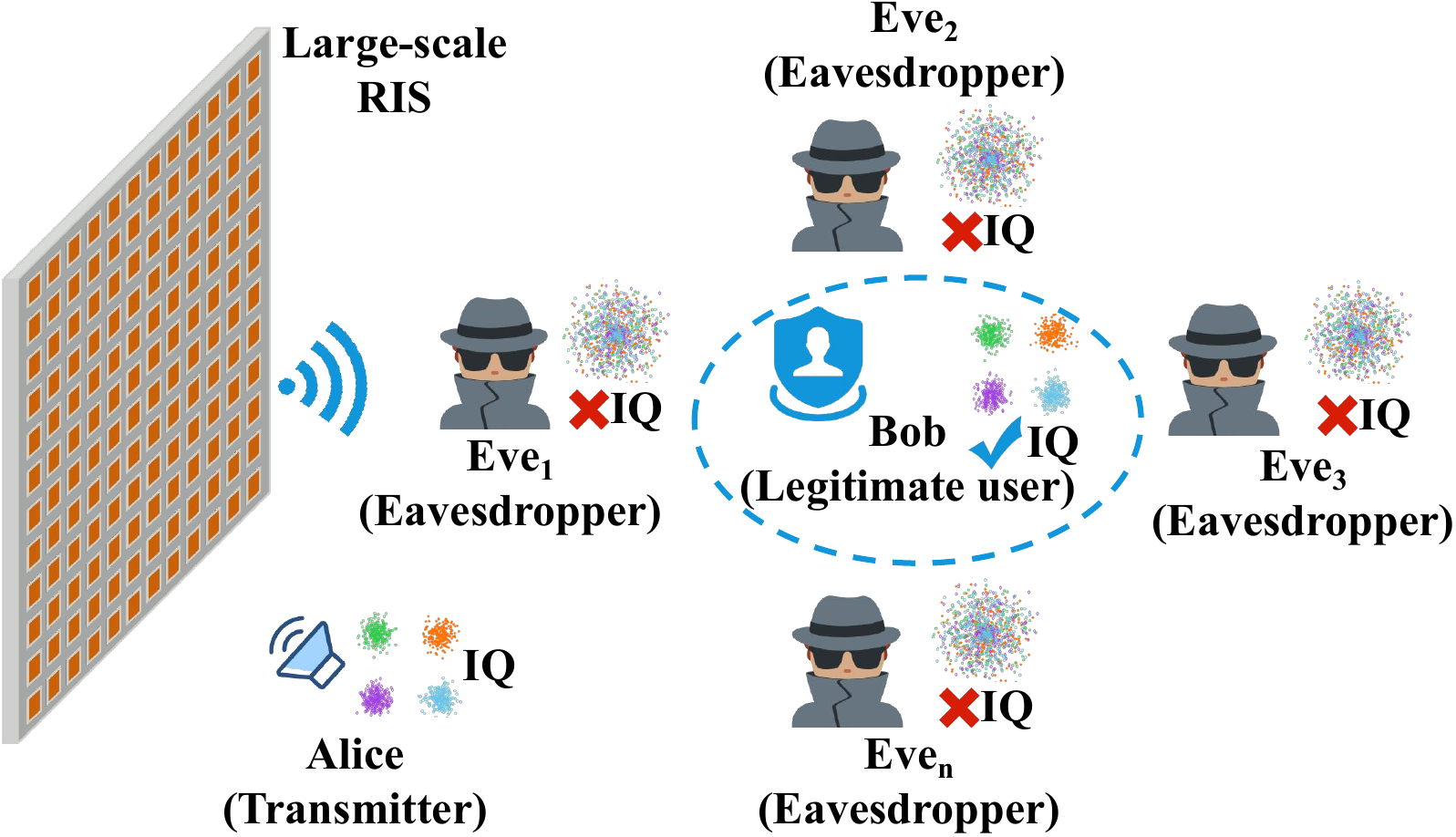}
\caption{The scenario of NF-SecRIS. The blue dashed line denotes the boundary of the secure communication region of Bob, within which reliable communication is maintained.}
\label{fig_01_scenario}
\vspace{0em}
\end{figure}%

Recent studies on RIS-assisted secure near-field communication have conducted modeling and simulating for the scenarios such as NOMA~\cite{10902048}, OAM~\cite{10910063}, and UAV~\cite{11121378}. Despite these advances, these works commonly assume a single eavesdropper with known location information. However, in practice, transmitters often lack prior knowledge of the number and position of eavesdroppers. Therefore, it is necessary to investigate 2D PLS in the presence of multiple potential eavesdroppers with unknown locations. Moreover, to the best of our knowledge, most related works remain at the simulation stage, with no prototype system developed for experimental validation \cite{10902048,10910063,11121378}. Furthermore, practical deployment requires security scheme with low complexity and real-time adaptability to prevent eavesdroppers from learning and adapting to fixed jamming patterns. To address these research gaps, we carry out further exploration in this work.

In this paper, \textbf{we propose NF-SecRIS, to our knowledge, the first large-scale RIS-based system for 2D PLS in near-field communications}.  Fig.~\ref{fig_01_scenario} depicts a representative scenario of NF-SecRIS. Modulated signals transmitted by Alice are reflected by the RIS, where only the legitimate user Bob in the near-field region can receive raw constellations. At other locations, whether differing in range or angle, eavesdroppers (Eves) can only obtain obfuscated constellations, thereby establishing 2D PLS in near-field communications. It addresses the limitation of conventional far-field PLS which lacks the control ability in the range dimension.

To achieve 2D PLS communications, the core objective lies in designing the dynamic RIS configurations  to enhance the quality of constellation at Bob while obfuscating constellations at other locations. Crucially, to prevent eavesdroppers from learning interference patterns for brute-force cracking, NF-SecRIS must transcend reliance on predictable RIS configurations by dynamically generating randomized phase patterns in real-time. This necessitates that our approach exhibits ultra-low computational complexity and high real-time capability. To address these requirements, we propose a \textbf{secure location modulation (SLM)} scheme with ultra-low complexity. Instead of brute-force optimization of the spatial-temporal phase patterns, our approach decouples the problem into spatial phase matrix synthesis and temporal phase sequence design, the latter further decomposed into basic sequence generation and interleaving sequence integration. Specifically, the framework of SLM comprises three modules: \textbf{(i) beam focusing with constant phase sequence; (ii) beam nulling with perturbed phase sequence; (iii) time-domain phase interleaving}.

Firstly, to enhance the signal strength at Bob while suppressing it at other locations, we leverage phase compensation principle to achieve near-field beam focusing at Bob's location. Simultaneously, time-domain constant phase sequence is employed to maintain constellation stability at Bob. However, although the signal strengths at eavesdroppers are suppressed, their constellations remain stable. Consequently, high-sensitivity eavesdroppers can effectively intercept raw physical layer signals. Therefore, it is essential to further obfuscate the constellations at eavesdroppers.

Secondly, to disrupt the constellations at Eves, we design an artificial noise generation scheme, i.e. beam nulling with perturbed phase sequence. To reduce the impact of artificial noise on Bob, the noise strength must generate a beam nulling at Bob's location. Yet, since the RIS is ultra-large-scale with nearly a thousand elements, direct optimization of spatial phase matrix faces prohibitively high complexity in a huge searching space. Hence, the primary challenge in generating beam nulling lies in reducing the computational complexity. To overcome this challenge, we propose a near-field squeeze-nulling model. By employing multi-focus squeezing to generate a null center, the optimization model dimensionality is reduced from nearly $10^{3}$ to merely 8, thus dramatically lowering the computational complexity.

Upon obtaining the spatial phase matrix for beam nulling, we face an another critical challenge: how to generate effective artificial noise in real time by leveraging beam nulling pattern. To address this challenge, we propose a low-complexity perturbed phase sequence scheme. Specifically, we develop a closed-form validation criterion for basic perturbed phase sequences, enabling rapid verification for the effectiveness of them. Concurrently, we construct a library of validated perturbed phase sequences which are pre-stored in the RIS controller for real-time invocation.

Thirdly, to simultaneously maintain the stability of constellation at Bob while disrupting the constellations at Eves, a tough challenge arises: how to integrate the beam focusing and beam nulling modules. Traditional method that directly integrates the spatial phase matrices will degrade near-field beam characteristics because of compromising the RIS effective aperture. Moreover, to prevent Eves from learning interference patterns for brute-force cracking, the integrating scheme must exhibit real-time randomness. To overcome this challenge, we propose a time-domain phase interleaving scheme that hybridizes perturbed sequence with constant sequence into interleaving sequence. Most critically, we develop closed-form lower and upper bounds for the ratio of time-slot allocation, drastically reducing the computational complexity. By randomly switching basic sequences and their time-slot ratios in real time, it ultimately generates dynamic configuration of RIS, achieving 2D PLS in near-field communications.

We fabricate a NF-SecRIS prototype featuring an ultra-large-scale RIS array of $14 \times 56$ elements operating at 5.8 GHz. Comprehensive experiments demonstrate that NF-SecRIS can achieve 2D PLS for diverse modulation schemes, such as ASK, FSK, PSK, and QAM, without requiring synchronization with transmitter or receiver. Although experimentally validated at 5.8 GHz, the NF-SecRIS framework is directly extensible to other frequency bands for 2D PLS.

In summary, the primary contributions of NF-SecRIS are as follows:

1) To our knowledge, NF-SecRIS is the first ultra-large-scale RIS-based system achieving 2D PLS in near-field communications. We propose a secure location modulation scheme that decouples the spatial-temporal phase patterns into spatial phase matrices and temporal phase sequences, dramatically reducing system design complexity.

2) We propose a low-complexity  squeeze-nulling model to synthesize the spatial phase matrix for beam nulling. By employing multi-focus squeezing to generate a null center, the optimization dimensionality is reduced from nearly $10^{3}$ to merely 8.

3) We propose a closed-form validation criterion for the effectiveness of basic perturbed phase sequences. Furthermore, we construct a library of validated sequences and pre-store them in the RIS controller for random real-time deployment.

4) We develop closed-form bounds for the time-slot ratios between cosntant and perturbed sequences, enabling rapid generation of interleaving sequences. This ultimately facilitates real-time stochastic synthesis of spatial-temporal phase patterns. Our experiments validate the feasibility and efficacy of NF-SecRIS, demonstrating its capability to achieve 2D PLS across diverse modulation schemes.





\section{System Overview of NF-SecRIS}

In this section, we introduce the design target and system architecture of NF-SecRIS.
\vspace{-1.2em}
\subsection{Design Target}

The overall objective of NF-SecRIS is to implement 2D PLS near-field communications. The scenario is shown in Fig.~\ref{fig_01_scenario}. Alice is the transmitter, while Bob is the legitimate receiver. There are several position-unknown eavesdroppers (Eves) surrounding Bob. In order to prevent Eves from eavesdropping on the communication data, an ultra-large-scale RIS is introduced in the communication link between Alice and Bob. Moreover, both Alice and Bob are located in the near-field region of the RIS. We assume that the RIS is aware of the locations of Alice and Bob, but remains unaware of the number and locations of Eves. There is no direct link between Alice and Bob. The signals transmitted by Alice are reflected by the RIS and then received by Bob and potential Eves. Only Bob can receive the raw constellations, while Eves at other ranges or angles can only receive the obfuscated constellations. It significantly differs from prior works \cite{10.1145/3495243.3560547} that failed to achieve secure communications in the range dimension. Our work is dedicated to the downlink transmission from Alice to Bob.


\vspace{-1.2em}
\subsection{System Architecture}

\begin{figure*}[!t]
\centering
\includegraphics[width=500pt]{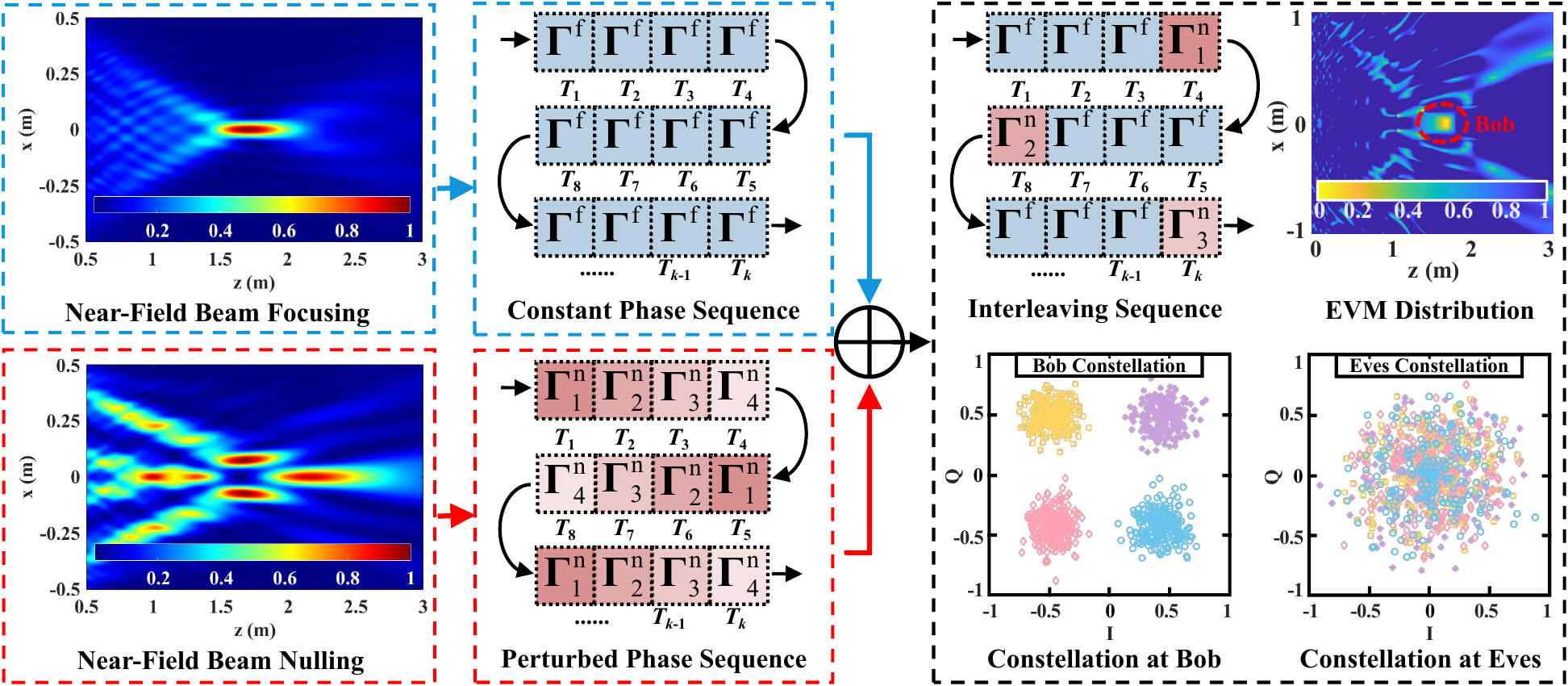}
\caption{System architecture of NF-SecRIS: Secure Location Modulation. The SLM scheme comprises three modules: (i) beam focusing with constant phase sequence (blue dashed box); (ii) beam nulling with perturbed phase sequence (red dashed box); (iii) time-domain phase interleaving (black dashed box).}
\label{fig_03_overview}
\vspace{-1.8em}
\end{figure*}

To achieve 2D PLS near-field communications, we propose a secure location modulation scheme to synthesize the near-field spatial-temporal phase patterns of the RIS, as illustrated in Fig.~\ref{fig_03_overview}. This method comprises three core modules: (i) beam focusing with constant phase sequence; (ii) beam nulling with perturbed phase sequence; (iii) time-domain phase interleaving. 

Firstly, to enhance Bob's signal strength while suppressing potential eavesdroppers, the RIS achieves near-field beam focusing via phase compensation principle. Simultaneously, to maintain constellation stability at Bob, we employ constant phase sequence in the time domain, whereby each RIS element maintains fixed phase shifts across all time slots. This module establishes a basis for 2D PLS near-field communications.

Secondly, to disrupt the constellations of Eves, we design an artificial noise scheme. This scheme first controls the spatial distribution of the noise energy via beam nulling, ensuring low noise levels at Bob's location and high levels in the surrounding areas. Subsequently, the artificial noise is generated in practice by using a perturbed phase sequence in the time domain.

Finally, to achieve 2D PLS communications, the focusing and nulling modules must be jointly operated. 
We propose a time-domain phase interleaving scheme. By randomly switching constant and perturbed sequences, the RIS generates real-time spatial-temporal phase pattern that prevents eavesdroppers from learning interference patterns for brute-force attacks, while maintaining reliable communications for Bob.


\section{Secure Location Modulation}

In this section, we elaborate on the secure location modulation secheme of NF-SecRIS. 

\vspace{-1.2em}
\subsection{Beam Focusing with Constant Phase Sequence}

To achieve 2D PLS, the fundamental objective is to enhance the signal strength at Bob while suppressing it at other locations. Furthermore, to ensure reliable communications toward Bob, it is essential to maintain the stability of its constellation. Therefore, we design a beam focusing module that computes the spatial phase matrix based on the principle of spherical-wave phase compensation, and employs a constant phase sequence in the time domain to guarantee the stability of constellation at Bob.

\begin{figure}[!t]
\centering
\subfloat[General model]{\includegraphics[height=120pt]{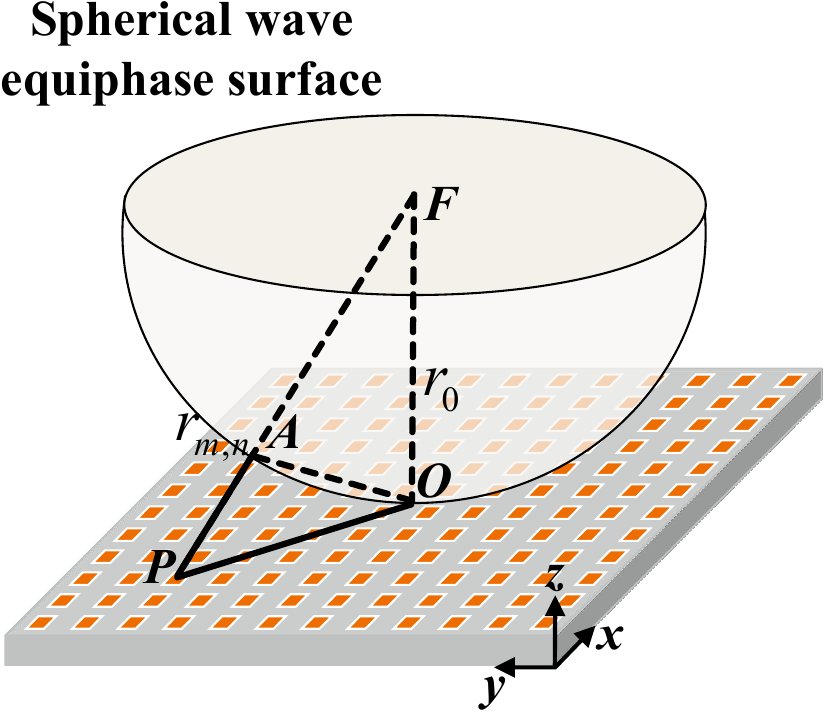}%
\label{fig_04a_聚焦通用模型}}
\hfil
\subfloat[RIS model]{\includegraphics[height=120pt]{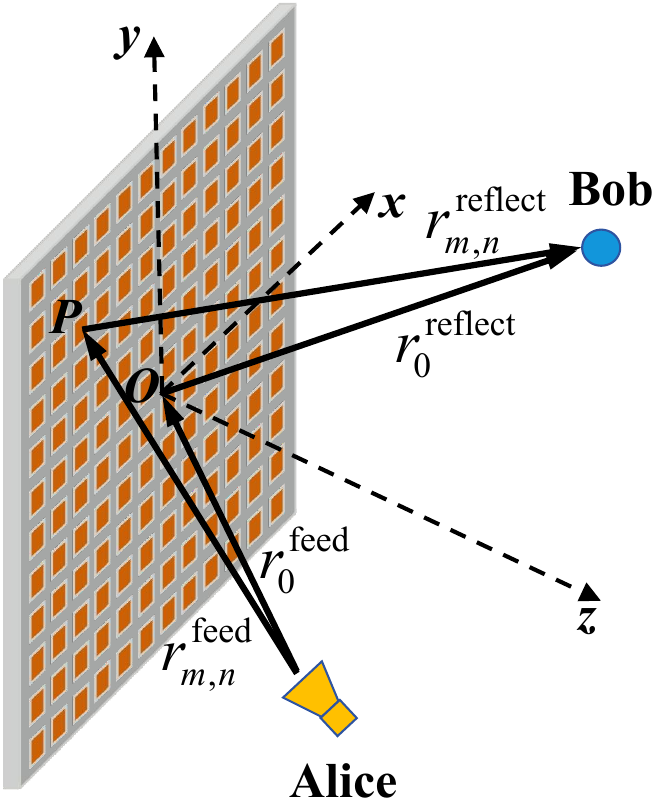}%
\label{fig_04b_聚焦RIS模型}}
\caption{Near-field phase compensation principle for beam focusing. (a) General spherical wave model. (b) Beam focusing model of RIS.}
\label{fig_04_focus_model}
\vspace{-1.8em}
\end{figure}

First, we synthesize the spatial phase matrix for beam focusing via the near-field spherical-wave phase compensation principle. As illustrated in Fig.~\ref{fig_04a_聚焦通用模型}, the planar array consists of $M \times N$ unit cells. Here, $O$ denotes the array center, $F$ represents the focal point, and $P$ indicates the center coordinates of the $(m, n)^{th}$ unit cell, where $m \in [1,M]$ and $n \in [1, N]$. Point $A$ is the intersection of the line segment $\overline{PF}$ and the equiphase surface of the spherical wave. The distance $|\overline{FO}|$ serves as the reference propagation distance, satisfying $|\overline{FO}|=|\overline{FA}|=r_{0}$. The propagation distance from the $(m, n)^{th}$ unit cell to the focal point $F$ is given by $|\overline{PF}|=r_{m,n}$. Therefore, the phase compensation for this unit cell is calculated as $\phi_{m,n}=\frac{2\pi}{\lambda}(r_{m,n}-r_{0})$, where $\lambda$ is the wavelength.

As shown in Fig.~\ref{fig_04b_聚焦RIS模型}, the aforementioned phase compensation model is extended to the scenario of RIS. The signals emitted by Alice are reflected by the RIS and then propagate as a spherical wave to Bob. The total reference distance is $r_0=r_{0}^{\text{feed}}+r_{0}^{\text{reflect}}$, where $r_{0}^{\text{feed}}$ and $r_{0}^{\text{reflect}}$ represent the distances from Alice and Bob to the center of the array, respectively. The total propagation distance for the $(m, n)^{th}$ unit cell is $r_{m,n}=r_{m,n}^{\text{feed}}+r_{m,n}^{\text{reflect}}$, in which $r_{m,n}^{\text{feed}}$ and $r_{m,n}^{\text{reflect}}$ denote the distance from Alice and Bob to this unit cell, respectively. Therefore, the ideal compensation phase is
%
%
%
%
%
%
%
%
\begin{align}
\phi_{m,n}^{\text{ideal}} 
 = \frac{2\pi}{\lambda}(r_{m,n}^{\text{feed}}+r_{m,n}^{\text{reflect}}-r_{0}^{\text{feed}}-r_{0}^{\text{reflect}}) .
 \label{eq_02_focus}
\end{align}
%
%
%

The NF-SecRIS system employs a 2-bit RIS, where the reflection amplitude of each unit cell is fixed at 1, while only four discrete phase shifts are available: $0$, $\pi/2$, $\pi$, and $3\pi/2$. Therefore, the ideal compensation phase must be quantized. The resulting quantized focusing phase $\phi_{m,n}^{\text{focus}}$ yields a reflection coefficient $\Gamma_{m,n}^{\text{focus}}$, which can be expressed as
%
%
%
%
%
%
%
%
\begin{align}
\Gamma_{m,n}^{\text{focus}} 
&= Q\left( \Gamma_{m,n}^{\text{ideal}} \biggm|_{2-bit} \right)
= Q\left(e^{j\phi_{m,n}^{\text{ideal}}} \biggm|_{2-bit} \right) \notag\\
&= \begin{cases}
    1, & -\pi/4 \leq \phi_{m,n}^{\text{ideal}} < \pi/4   \\
    j, & \pi/4 \leq \phi_{m,n}^{\text{ideal}} < 3\pi/4   \\
    -1,  & 3\pi/4 \leq \phi_{m,n}^{\text{ideal}} < 5\pi/4  \\
    -j,  & \text{otherwise}   
   \end{cases}.
   \label{equ_量化规则}
\end{align}
%
%
In all subsequent sections of this paper, the phase of RIS defaults to the quantized phase.

\begin{figure}[!t]
\centering
\subfloat[2D beam pattern]{\includegraphics[height=84pt]{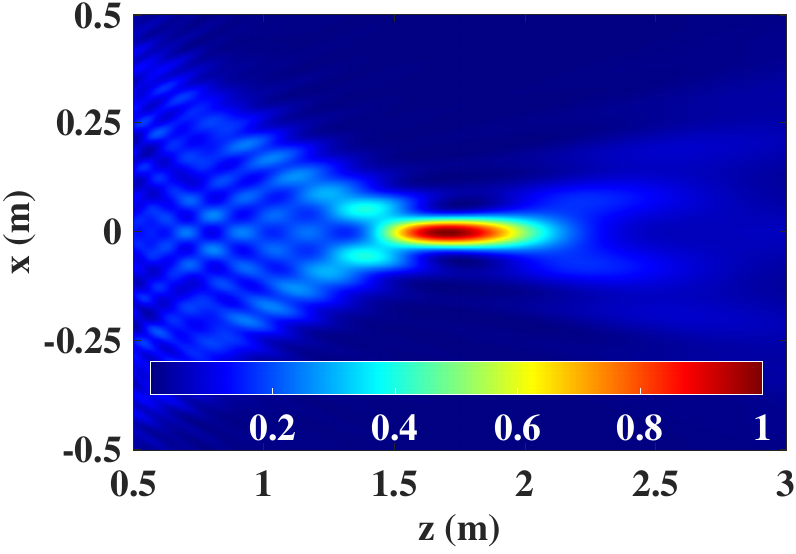}%
\label{fig_05b_聚焦二维}}
\hfil
\subfloat[Beam pattern vs. $r$]{\includegraphics[height=84pt]{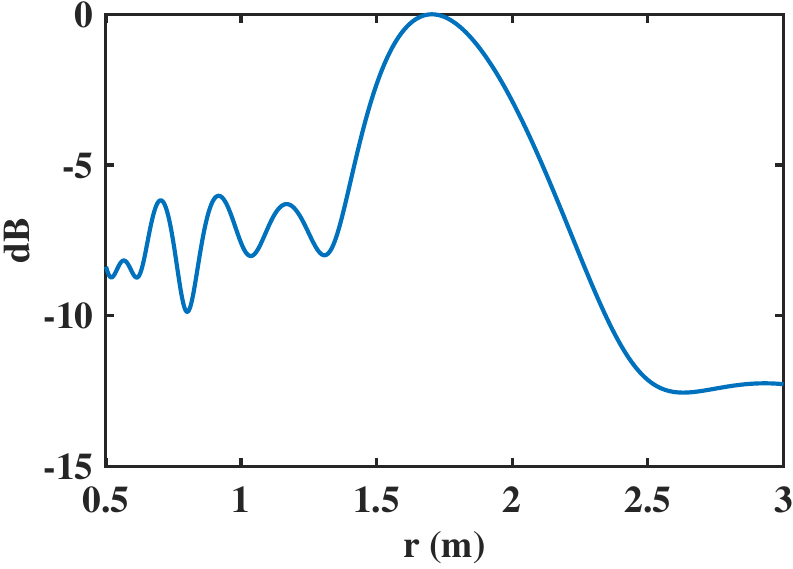}%
\label{fig_05c_聚焦距离}}
\caption{Near-field beam focusing. (a) Power distribution in $xoz$ plane. (b) Power distribution versus distance $r$. }
\label{fig_05总_聚焦空间编码}
\vspace{-1.8em}
\end{figure}

Within the same time slot, the quantized phase distribution across all unit cells of the RIS array is termed as the spatial phase matrix. The spatial phase matrix for beam focusing is denoted as $\bm{\Phi}^{\text{f}} = [\phi_{m,n}^{\text{focus}}] \in \mathbb{R}^{M \times N}$. As observed in Fig.~\ref{fig_05b_聚焦二维}, beam focusing is achieved in the $xoz$ plane. Fig.~\ref{fig_05c_聚焦距离} provides an unfolded view of Fig.~\ref{fig_05b_聚焦二维} along the distance $r$, clearly demonstrating that the near-field beam forms a distinct main lobe in the range dimension, rather than exhibiting monotonic decay with increasing distance as in the far-field case. This represents a fundamental distinction between near-field focusing and conventional far-field beamforming.

Since the reflection state of each RIS unit can change instantaneously over time, the spatial phase matrix can differ from one time slot to another. This implies that the RIS can generate a spatial-temporal phase patterns over a period. Directly optimizing the dynamic phase patterns would lead to prohibitively high complexity due to the enormous search space. To reduce the complexity, we introduce a time-domain phase sequence on top of a static spatial phase matrix to synthesize the spatial-temporal phase patterns. This sequence describes the unified variation pattern of each unit in the spatial phase matrix along the time dimension. For example, if the value of the sequence in a certain time slot is $\pi/2$, it means that an additional phase shift of $\pi/2$ is applied to every unit of the spatial phase matrix in that slot. Under this operation, the relative phase differences among units remain unchanged, so the spatial energy distribution is unchanged; however, the absolute phase is altered, resulting in dynamic rotation of the constellation. This lays the foundation for subsequent differentiated manipulation of the constellation quality for Bob and Eves. In the following, we provide a detailed analysis of the time-domain phase sequence.

Without loss of generality, we assume that the time-domain pulse width of the RIS is $\tau$. Within a single pulse, the state of the RIS remains constant, while reconfiguration is permitted across different pulse slots. The time-varying behavior of the beam-focusing reflection coefficient matrix $\bm{\Gamma}^{\text{f}} \in \mathbb{C}^{M \times N}$ can be expressed as 
%
%
%
%
%
%
%
%
\begin{align}
    \bm{\Gamma}^{\text{f}}(t)
    &= \sum_{k=1}^{K}\bm{\Gamma}_{k}^{\text{f}}g(t-(k-1)\tau) \notag\\
    &= \sum_{k=1}^{K}\mathbf{A}_{k}^{\text{f}} \odot 
        e^{j\bm{\Phi}_{k}^{\text{f}}}
        g(t-(k-1)\tau) ,
\end{align}%
where the system operates over $K$ time slots, the reflection coefficient matrix for the $k$-th slot is $\bm{\Gamma}_{k}^{\text{f}}$. Specifically, $\bm{\Gamma}_{k}^{\text{f}}$ can be decomposed into the Hadamard product (element-wise multiplication, $\odot$) of a real-valued amplitude matrix $\mathbf{A}^{\text{f}}_{k} \in \mathbb{R}^{M \times N}$ and a complex-valued phase matrix $e^{j\bm{\Phi}_{k}^{\text{f}}} \in \mathbb{C}^{M \times N}$. The temporal window function $g(t)$ is also defined as
%
%
%
%
%
%
%
%
\begin{align}
    g(t) = 
    \begin{cases}
        1, &  0 \leq t < \tau \\
        0, & \text{otherwise}
    \end{cases}.
\end{align}%

To simplify the complexity, the reflection amplitude of all unit cells is fixed at 1, which is mathematically equivalent to $\mathbf{A}^{\text{f}}_{k} \equiv \mathbf{J} \in \mathbb{R}^{M \times N}$, where $\mathbf{J}$ is an all-ones matrix, and the symbol $\equiv$ denotes “identically equal to”. Therefore, $\bm{\Gamma}^{\text{f}}(t)$ can be simplified as 
%
%
%
%
%
%
%
%
\begin{align}
    \bm{\Gamma}^{\text{f}}(t)
    &= \sum_{k=1}^{K} 
        e^{j\bm{\Phi}_{k}^{\text{f}}}
        g(t-(k-1)\tau) \notag\\
    &=  e^{j\bm{\Phi}^{\text{f}}}
        \sum_{k=1}^{K}
        R_{k}^{\text{f}}
        g(t-(k-1)\tau),
    \label{equ_信号时间编码2}
\end{align}
%
%
where $\bm{\Phi}^{\text{f}}$ and $R_{k}^{\text{f}}$ is spatial phase matrix and time-domain phase sequence for beam-focusing pattern, respectively. In the case of 2-bit RIS, the $R_{k}^{\text{f}}$ takes discrete values in $R_{k}^{\text{f}} \in \left\{1, j, -1, -j \right\} $. Equation (\ref{equ_信号时间编码2}) shows that introducing a time-domain phase sequence to the spatial phase matrix enables concurrent control of the energy distribution in space and the constellation rotation in the complex plane. 

\begin{figure}[!t]
    \centering
    \subfloat[Constant phase sequence in time domain]{\includegraphics[width=248pt]{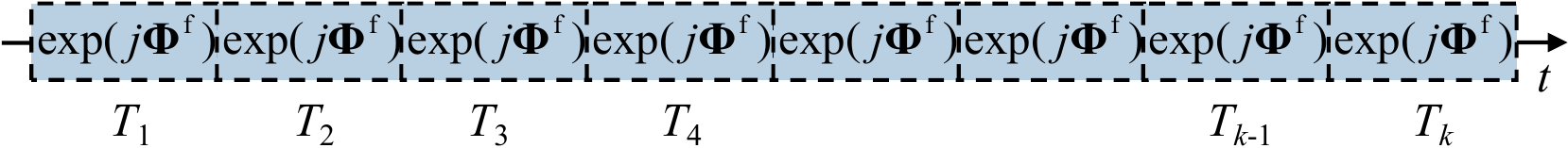}}
    \label{fig_06a_信号时序图} \\
    \subfloat[Alice]{\includegraphics[height=76pt]{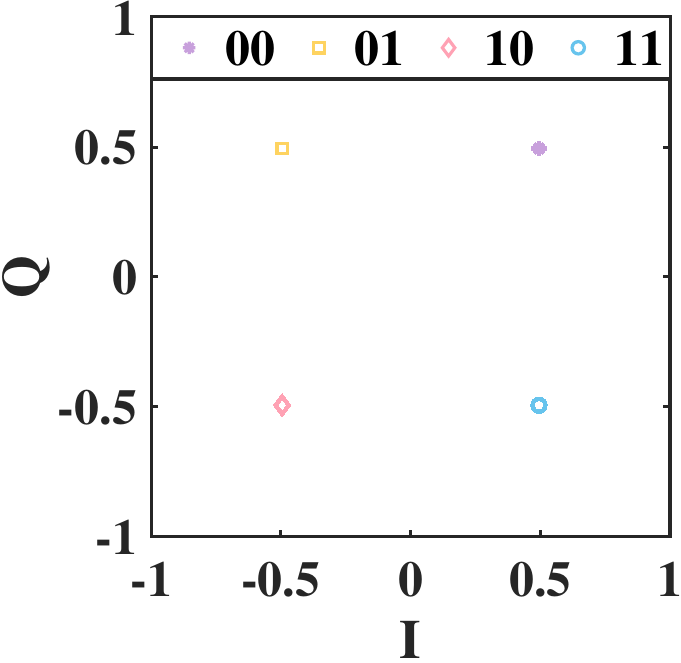}}
    \label{fig_06b_信号alice星座图}
    \hfil
    \subfloat[Bob]{\includegraphics[height=76pt]{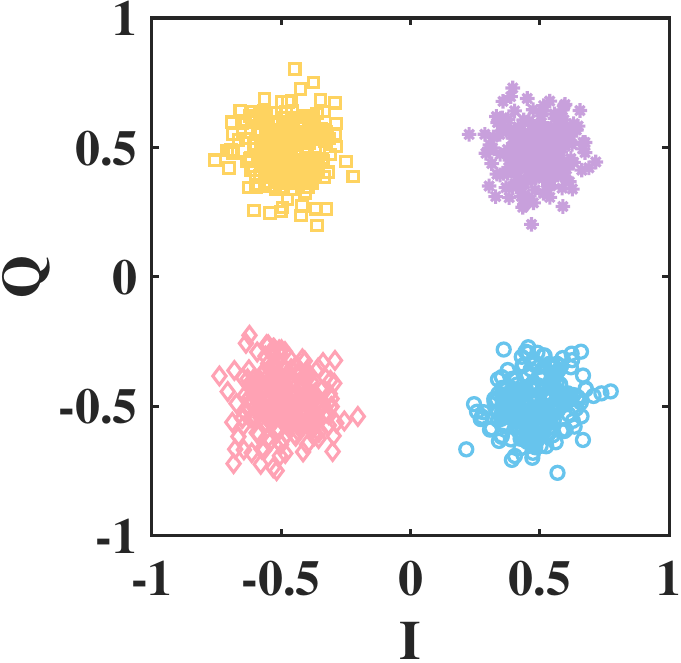}}
    \label{fig_06c_信号bob星座图}
    \hfil
    \subfloat[Eve]{\includegraphics[height=76pt]{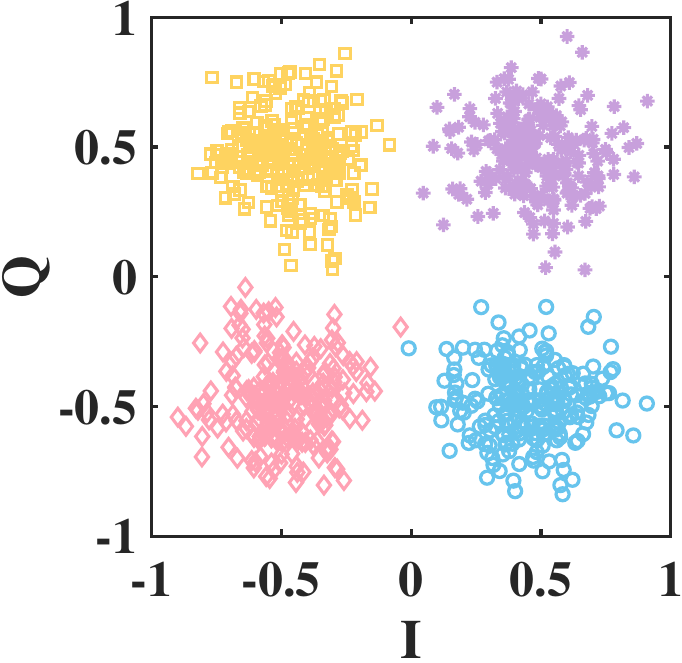}}
    \label{fig_06d_信号eve星座图}
    \caption{The influence of constant phase sequence on constellations. (a) Constant phase sequence in time domain. (b) Ideal constellation at Alice. (c) Constellation at Bob. (d) Constellation at Eve.}
    \label{fig_06总_信号时间编码}
    \vspace{-1.8em}
\end{figure}%

To ensure the stability of constellation at Bob, we set $R_{k}^{\text{f}} \equiv 1$, which leads to the result $\bm{\Gamma}^{\text{f}}(t) = e^{j\bm{\Phi}^{\text{f}}}$. Under this constant phase sequence, as illustrated in Fig. \ref{fig_06总_信号时间编码}, the time-domain phase sequence remains constant across all time slots. As a result, Bob, benefiting from high signal strength, can demodulate the constellation transmitted by Alice with high stability and clarity. In contrast, Eve suffers from a substantial degradation of constellation quality due to low signal strength.

\vspace{-1.2em}
\subsection{Beam Nulling with Perturbed Phase Sequence}

Although the beam focusing module suppresses the signal strengths of eavesdroppers, their constellations remain stable, allowing those with sufficiently high sensitivity to successfully intercept raw IQ data. To obfuscate constellations at Eves, we design a beam nulling module. This module synthesizes the spatial phase matrix for beam nulling via a squeeze-nulling model with low complexity and generates artificial noise using basic perturbed phase sequences.

The first challenge in generating artificial noise lies in determining its spatial energy distribution. For Bob, the noise intensity should be minimized to reduce interference. For Eves, however, artificial noise represents a limited interference resource, it need not be uniformly distributed in space. Instead, higher noise strength should be allocated to potential eavesdropping locations with strong signal levels, while lower noise strength suffices in regions with weaker signals. This strategy ensures that the signal-to-noise ratio (SNR) is effectively suppressed at all locations of Eves using finite noise resources. Since beam focusing module results in the highest signal strength around Bob, the noise energy should primarily surround Bob’s vicinity while being significantly suppressed exactly at its location, i.e. beam nulling. Beam nulling refers to a sharp field strength reduction at a specific location, while maintaining high field strength in the surrounding area. This technique forces the artificial noise to encircle Bob's location, while leaving almost no noise at Bob’s exact location.

However, unlike beam focusing, there is no explicit closed-form solution for beam nulling synthesis. Instead, optimization methods are required to solve the beam nulling problem. Specifically, assuming the reflection coefficient of each RIS unit cell is $\Gamma_{m,n}$, the normalized electric field intensity at a spatial point $\bm{r}$ can be expressed based on the ray-optics model in Fig.~\ref{fig_04_focus_model} as
%
%
%
%
%
%
%
%
\begin{align}
    E(\bm{r}) = \sum_{m=1}^{M}\sum_{n=1}^{N}
    \frac{\Gamma_{m,n}}{r_{m,n}^{\text{feed}}\cdot r_{m,n}^{\text{reflect}}}
    e^{-j\frac{2\pi}{\lambda}(r_{m,n}^{\text{feed}}+r_{m,n}^{\text{reflect}})}.
    \label{equ_通用场分析}
\end{align}%
According to Eq.~(\ref{equ_通用场分析}), the most direct method for beam nulling synthesis is to optimize the coefficients $\Gamma_{m,n}$ of all units. Nevertheless, for ultra-large-scale RIS, this becomes a high-dimensional optimization problem that is challenging to solve. For example, the RIS used in NF-SecRIS has $ 14 \times 56=784 $ unit cells, meaning the searching space is 784 dimensions. Therefore, developing an optimization model with lower complexity is necessary for beam nulling. 

To address this challenge, we propose a near-field squeeze-nulling model (SNM). In antenna radiation patterns, a deep null usually forms between two adjacent high-gain lobes, implying that a null at a specific location can be synthesized by effectively squeezing the surrounding high-gain beams. Inspired by this phenomenon, the core idea of SNM is to strategically place multiple high-gain focal points around the intended nulling location. The reflected wave from RIS to these focal points destructively interferes at the target location, forming a deep beam nulling. The phase matrix for each focal point is computed using Eq.~\ref{eq_02_focus}. The final nulling reflection coefficients are obtained through weighted summation of the reflection coefficients from all focal points, which significantly reduces the computational complexity.

\begin{figure}[!t]
    \centering
    \includegraphics[width=250pt]{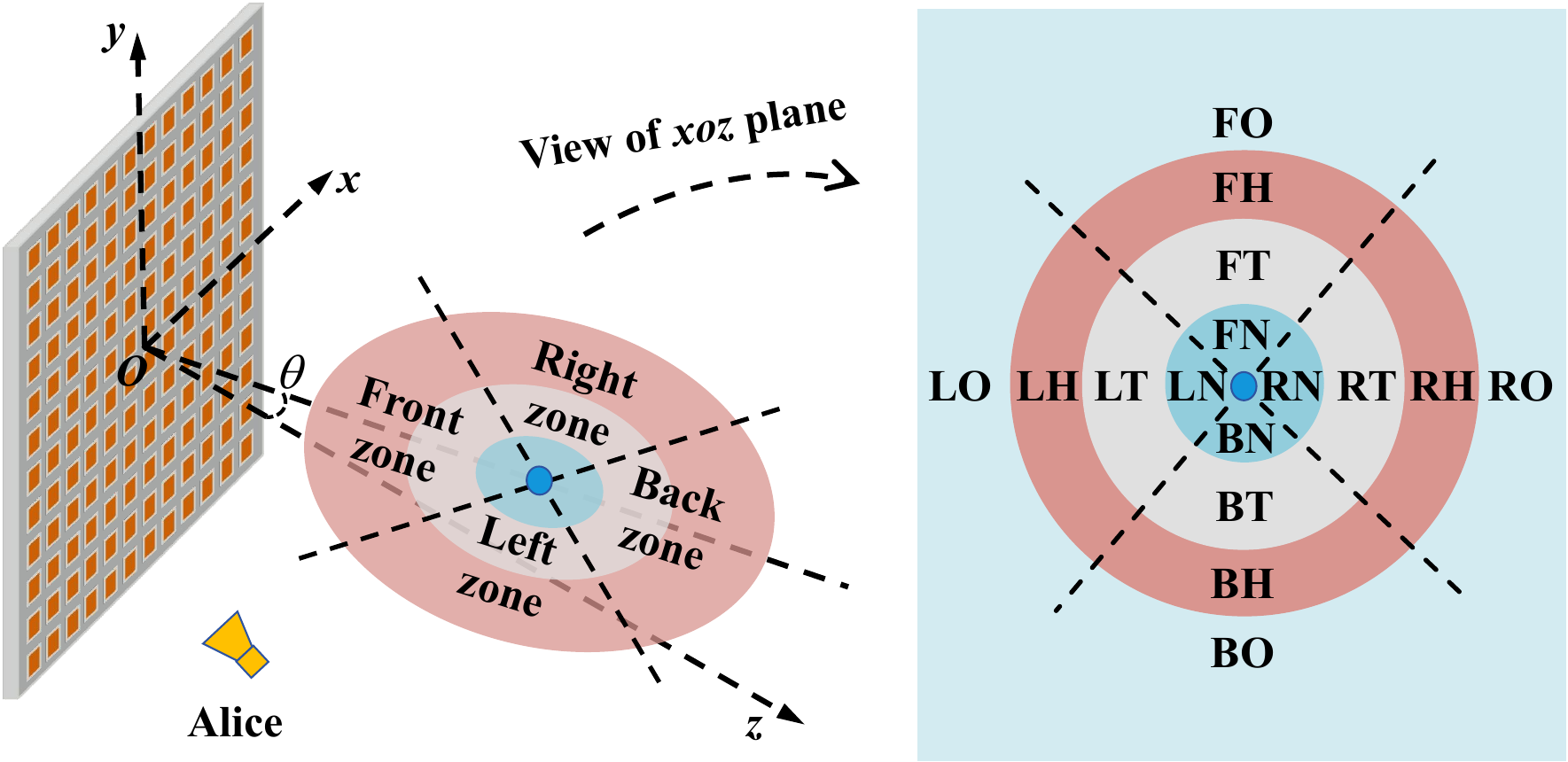}
    \caption{Near-field beam-nulling model. The meanings of the abbreviated letters are as follows: F (Front), B (Back), L (Left), R (Right), N (Nulling), T (Transition), H (High-gain), O (Outer). For example, FN represents the Front-Nulling zone, BH represents the Back-High-Gain zone.}
    \label{fig_07_零陷模型}
    \vspace{-1.8em}
\end{figure}%

\begin{figure}[!t]
    \centering
    \subfloat[2D beam pattern]{\includegraphics[height=84pt]{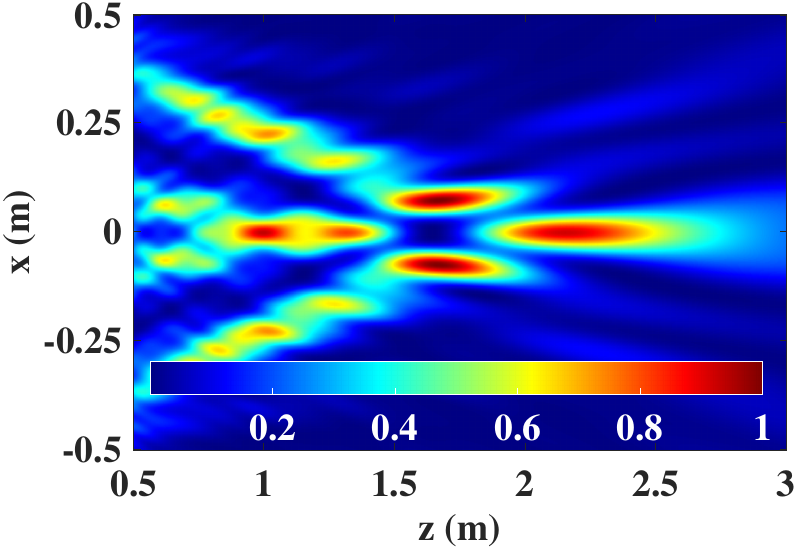}}
    \label{fig_08b_零陷二维图}
    \hfil
    \subfloat[Beam pattern vs. $r$]{\includegraphics[height=84pt]{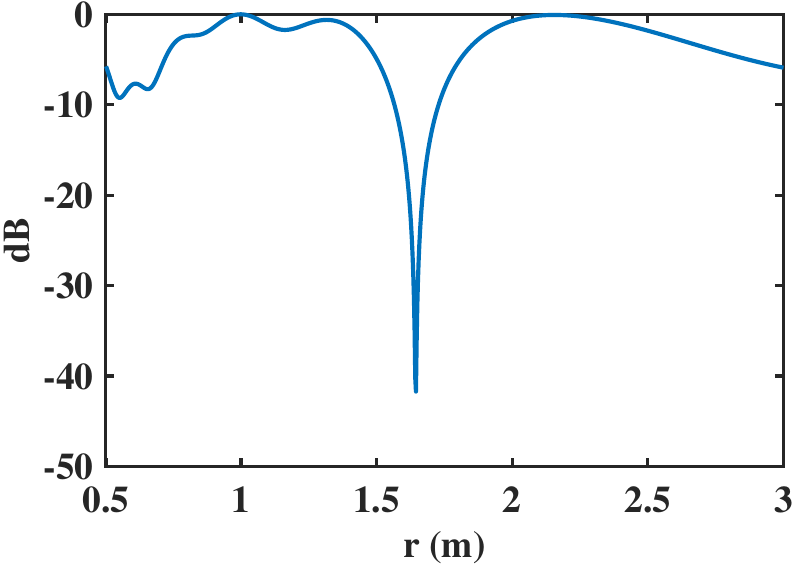}}
    \label{fig_08c_零陷一维图}
    \caption{Near-field beam nulling. (a) Power distribution in $xoz$ plane. (b) Power distribution versus distance $r$.}
    \label{fig_08总_零陷效果}
    \vspace{-1.8em}
\end{figure}%

The SNM is presented in Fig.~\ref{fig_07_零陷模型}. The region surrounding the null center is divided into four main zones: front zone, back zone, left zone, and right zone. A focal point is set in each zone, and through the squeezing of these four focal points, a deep null is formed at the target center. Each main zone is further subdivided into four subzones: nulling subzone, transition subzone, high-gain subzone, and outer subzone. It is worth noting that a transition subzone is introduced between the nulling and high-gain subzones, where the field strength varies rapidly and is not strictly constrained. Assuming the spherical coordinate of the null center is $\bm{p}_c=(r_{c},\theta_{c},\phi_{c})$, the position offset vector of the four focal points is defined as $\bm{d}=[\Delta r_{q_{f}}, \Delta r_{q_{b}}, \Delta \theta_{q_{l}}, \Delta \theta_{q_{r}}] \in \mathbb{R}^{4}$, where the subscripts $q_{f}, q_{b}, q_{l}, q_{r}$ denote the zone of $front, back, left, right$, respectively. Thus, the positions of the four focal points are given by
%
%
%
%
%
%
%
%
\begin{align}
    \bm{p}_{q} = 
    \begin{cases}
        (r_c - \Delta r_{q_{f}}, \theta_c, \phi_c), 
        & q \in \left\{q_{f} \right\}  \\
        (r_c + \Delta r_{q_{b}}, \theta_c, \phi_c), 
        & q \in \left\{q_{b} \right\}  \\        
        (r_c, \theta_c - \Delta \theta_{q_{l}}, \phi_c),
        & q \in \left\{q_{l} \right\} \\        
        (r_c, \theta_c + \Delta \theta_{q_{r}}, \phi_c),
        & q \in \left\{q_{r} \right\}
    \end{cases},
    \label{equ_夹逼焦点位置}
\end{align}%
where $\phi_c = 0$, since our analysis is confined to the beam nulling in the $xoz$ plane. Assuming the weight vector for the four focal points is $\bm{w}=[w_{q_{f}},w_{q_{b}},w_{q_{l}},w_{q_{r}}] \in \mathbb{R}^4$, the reflection coefficient $\Gamma_{m,n}$ in Eq.~(\ref{equ_通用场分析}) can be expressed as
%
%
%
%
%
%
%
%
\begin{align}
    \Gamma_{m,n} (\bm{d},\bm{w})
    &= Q \left(\sum_{q \in \left\{q_{f},q_{b},q_{l},q_{r} \right\} } w_{q} \cdot e^{j \phi_{m,n,q}} \biggm|_{2-bit} \right),
    \label{equ_零陷相位合成及量化}
\end{align}%
where $Q$ denotes the quantization function, consistent with the rule specified in Eq.~(\ref{equ_量化规则}). The phase $\phi_{m,n,q}$ for each focal point, based on the phase compensation principle in Eq.~\ref{eq_02_focus}, is formulated as
%
%
%
%
%
%
%
%
\begin{align}
    \phi_{m,n,q} 
    &= \frac{2\pi}{\lambda} \left(r_{m,n}^{\text{feed}}+r_{m,n}^{\text{reflect}}(\bm{p}_{q})-r_{0}^{\text{feed}}-r_{0}^{\text{reflect}}(\bm{p}_{q})\right) .
    \label{equ_零陷夹逼理想相位}
\end{align}%

\begin{algorithm}[!t]
\caption{Near-Field Squeeze-Nulling Model}
\label{algo_01_null}
\begin{algorithmic}[1]
\REQUIRE
Wavelength $\lambda$, array scale $M,N$, unit distance $\Delta x, \Delta y$, feed source position $\bm{r}^{\text{feed}}$, null center $\bm{r}_{c}$, sampling points $\bm{r}_{q,i}$
\ENSURE
Focus offset vector $\bm{d}^{\ast}$, weight vector $\bm{w}^{\ast}$, reflection coefficient matrix $\bm{\Gamma}^{\text{null}}$
\STATE \textbf{Initialization:} $k=0$, $\bm{d}^{(0)}$, $\bm{w}^{(0)}$; \\
\STATE \textbf{Fitness Analysis:}
\STATE Calculate $\bm{p}_{q}$ of 4 squeezed focus points via (\ref{equ_夹逼焦点位置}); \\
\STATE Calculate ideal phase $\phi_{m,n,q}$ via (\ref{equ_零陷夹逼理想相位}); \\
\STATE Synthesize reflection coefficients $\Gamma_{m,n}^{\text{null}}$ via (\ref{equ_零陷相位合成及量化}); \\
\STATE Analyze $E(\bm{r}_{q,i})$ of sampling points via (\ref{equ_通用场分析}); \\
\STATE Calculate null depth via (\ref{equ_零陷优化模型}) and (\ref{equ_场强均值}); \\
\STATE \textbf{Optimization Loop:} \\
\WHILE{Not converged and $k \leq K_{\max}$}
\STATE Update $\bm{d}^{(k)}$ and $\bm{w}^{(k)}$ using specific optimization algorithm, e.g. GA; \\
\STATE \textbf{Fitness Analysis} as mentioned above; \\
\STATE $k=k+1$; \\
\ENDWHILE \\
\RETURN $\bm{d}^{\ast}$, $\bm{w}^{\ast}$,  $\bm{\Gamma}^{\text{null}}$.
\end{algorithmic}
\end{algorithm}%

The core task now is that determining the offset vector $\bm{d}$ and the weight vector $\bm{w}$ to obtain the reflection coefficient $\Gamma_{m,n}$. We use nulling depth as the key metric, defined as the ratio  of the average field strength amplitude in the nulling subzone to that in the high-gain subzone. The optimization model is formulated as 
%
%
%
%
%
%
%
%
%
%
%
%
\begin{align}
    &\min_{\bm{d},\bm{w}} \sum_{q \in \left\{q_{f},q_{b},q_{l},q_{r} \right\}}
    \frac{\overline{\left| E_{q}^{null} \right|}}{\overline{\left| E_{q}^{high} \right|}} \notag \\
    & \text{s.t.} \notag \\
    & \text{C1:} \, w_{q} \geq 0, \sum_{q \in \left \{ q_{f},q_{b},q_{l},q_{r} \right\}}
    w_{q} = 1, \notag \\
    & \text{C2:} \, \frac{\max \left| E_{q}^{high}\right|}{\max \left| E_{q}^{outer}\right|} \geq 1, \, \forall q \in \left\{ q_{f},q_{b},q_{l},q_{r} \right\}, \notag \\
    & \text{C3:} \, \frac{\max \left| E_{q_{1}}^{high}\right|}{\max \left| E_{q_{2}}^{high}\right|} \in [\alpha_{1},\alpha_{2}], \, \forall q_{1},q_{2} \in \left\{ q_{f},q_{b},q_{l},q_{r} \right\},
    \label{equ_零陷优化模型}
\end{align}%
where
%
%
%
%
%
%
%
%
\begin{align}
    \frac{\overline{\left| E_{q}^{null} \right|}}{\overline{\left| E_{q}^{high} \right|}}
    = \frac{\frac{1}{N_{q}^{null}} \sum_{i=1}^{N_{q}^{null}} \left| E_{q}^{null} (\bm{r}_{i}) \right|}{\frac{1}{N_{q}^{high}} \sum_{j=1}^{N_{q}^{high}} \left| E_{q}^{high} (\bm{r}_{j}) \right|} .
   \label{equ_场强均值}
\end{align}%
Here, $N_{q}^{null}$ and $N_{q}^{high}$ denotes the number of sampling points for nulling subzone and high-gain subzone, respectively. The ratio of the average field strength in the nulling subzone to that in the high-gain subzone defines the nulling depth of that zone. The sum of the nulling depths across all four zones serves as the objective function, ensuring comprehensive nulling performance over the entire domain. Constraint C2 is a peak position constraint, which guarantees that the highest gain within each main zone lies in the high-gain subzone rather than in outer subzone. Constraint C3 is a peak balance constraint, ensuring that the peak gains among different main zones remain balanced, with $\alpha_{1}$ and $\alpha_{2}$ representing the lower and upper bounds of this constraint, respectively. Based on our simulation observations, we set $\alpha_{1} = 0.9 $ and $\alpha_{2} = 1.1$.
Specifically, for NF-SecRIS with 784 unit cells, the SNM reduces the dimension of searching space from 784 to only 8, significantly alleviating the computational complexity. We employ a genetic algorithm to solve SNM. The overall solution process is summarized in Algorithm~\ref{algo_01_null}.

The simulation results of beam nulling using SNM is shown in the Fig.~\ref{fig_08总_零陷效果}. Specifically, Fig.~\ref{fig_08总_零陷效果}a shows the simulated power distribution in the $xoz$ plane, and Fig.~\ref{fig_08总_零陷效果}b provides a one-dimensional unfolded view along the distance $r$ based on Fig.~\ref{fig_08总_零陷效果}a. The results clearly demonstrate the formation of a deep nulling at the target center, surrounded by four focal points, meeting the requirements for noise energy distribution.

\begin{figure}[!t]
    \centering
    \subfloat[Perturbed phase sequence in time domain]{\includegraphics[width = 248pt]{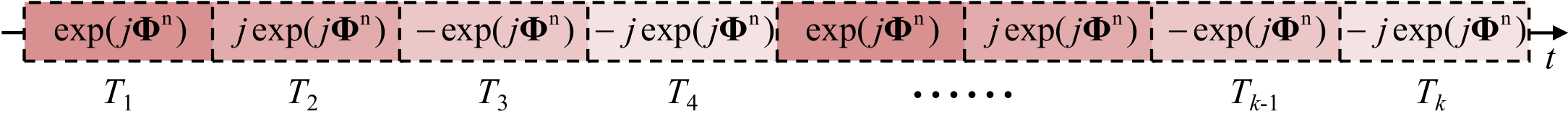}}
    \label{fig_09a_噪声时序图} \\
    \subfloat[$T_{1}~(0^\circ)$]{\includegraphics[height=55pt]{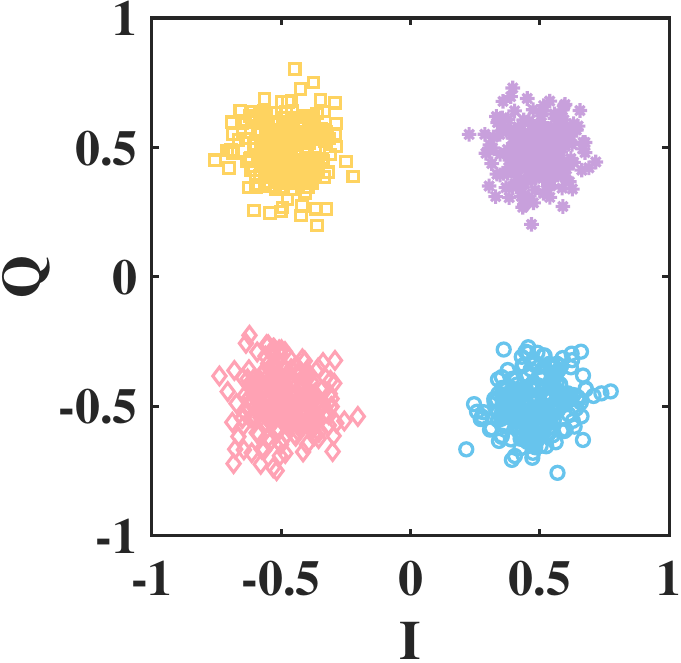}} 
    \label{fig_09b_0度} 
    \hfil
    \subfloat[$T_{2}~(90^\circ)$]{\includegraphics[height=55pt]{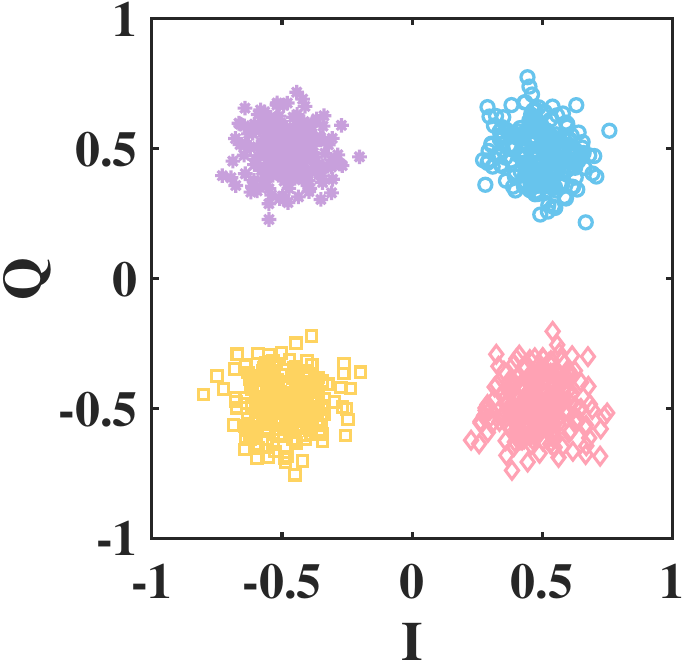}}
    \label{fig_09c_90度}
    \hfil
    \subfloat[$T_{3}~(180^\circ)$]{\includegraphics[height=55pt]{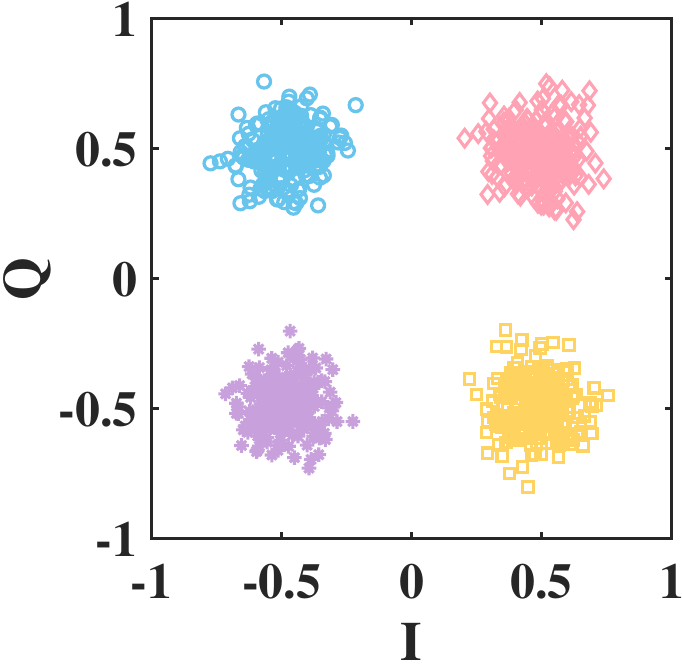}}
    \label{fig_09d_180度}
    \hfil
    \subfloat[$T_{4}~(270^\circ)$]{\includegraphics[height=55pt]{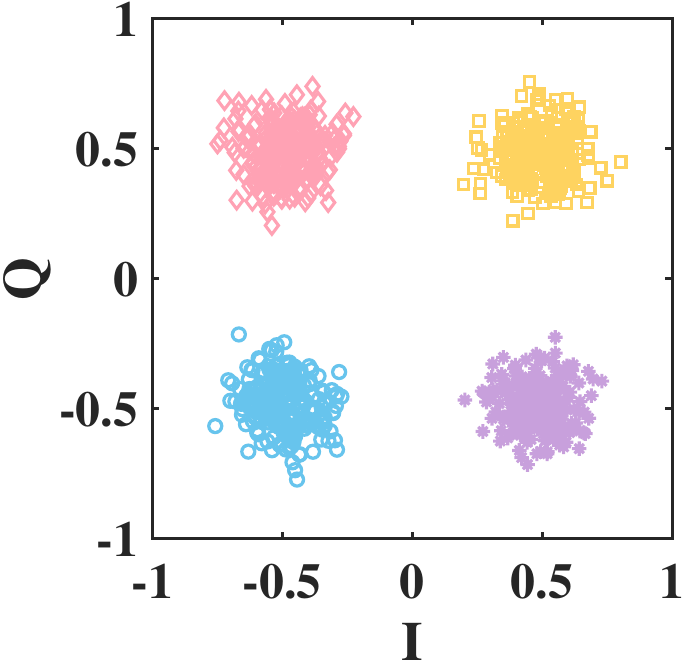}}
    \label{fig_09e_270度}
    \caption{The influence of perturbed phase sequence on constellations. (a) Perturbed phase sequence in time domain. (b) $R_{k}^{\text{n}}=1$. (c) $R_{k}^{\text{n}}=j$. (d) $R_{k}^{\text{n}}=-1$. (e) $R_{k}^{\text{n}}=-j$.}
    \label{fig_09总_噪声时间编码}
    \vspace{-1.8em}
\end{figure}

After obtaining the nulling phase matrix, a time-domain phase sequence is superimposed to induce dynamic rotation of the constellation in the complex plane, which is utilized for artificial noise generation. Similarly to the time-domain analysis denoted by Eq.~(\ref{equ_信号时间编码2}), the time-domain characteristics of the nulling module can be formulated as
%
%
%
%
%
%
%
%
\begin{align}
    \bm{\Gamma}^{\text{n}}(t)
    &=  e^{j\bm{\Phi}^{\text{n}}}
        \sum_{k=1}^{K}
        R_{k}^{\text{n}}
        g(t-(k-1)\tau),
    \label{equ_噪声时间编码}
\end{align}%
where $\bm{\Phi^{\text{n}}}$ is the nulling phase matrix, and $R_{k}^{\text{n}} \in \left\{1, j, -1, -j \right\} $ is the time-domain phase sequence. In contrast to the constant phase sequence used in the beam focusing module, the phase sequence of the nulling module must vary dynamically to generate noise, hence we introduce a perturbed phase sequence $R_{k}^{\text{n}}$. For 2-bit RIS, Fig.~\ref{fig_09总_噪声时间编码} illustrates the relationship between the perturbed phase and the rotation state of the constellation.

How to effectively utilize the four basic states of $R_{k}^{\text{n}}$ to generate potent artificial noise presents a significant challenge. Additionally, to prevent Eves from learning the interference pattern, the noise must be random and change in real-time. To meet these requirements, we establish a library of basic perturbed phase sequences. These pre-computed perturbed phase sequences are stored in the RIS controller, and artificial noise is generated by randomly calling upon them in real-time. Building such a vast library efficiently demands a low-complexity generation technique. Thus, we develop a closed-form validation criterion to quickly evaluate basic perturbed phase sequences, the exact expression of which will be derived in the next subsection.

\vspace{-1.2em}
\subsection{Time-Domain Phase Interleaving}

To simultaneously enhance Bob's signal strength while maintaining its constellation stability, and suppress Eves' signals while disrupting their constellations, the beam focusing and beam nulling modules must be integrated. Accordingly, we design a time-domain phase interleaving scheme, derive a closed-form validation criterion for basic perturbed phase sequences, construct a basic sequence library, and establish closed-form upper and lower bounds for the time-slot ratio allocation between perturbed and constant phase sequences.

We assume that the time-domain characteristics of interleaving sequence can be expressed as
%
%
%
%
%
%
%
%
\begin{align}
    \bm{\Gamma}^{\text{i}}(t)
    &= \sum_{k=1}^{K}\bm{\Gamma}_{k}^{\text{i}}g(t-(k-1)\tau),
    \label{equ_14_联合时空编码}
\end{align}%
where $\bm{\Gamma}_{k}^{\text{i}} \in \left\{e^{j\bm{\Phi^{\text{f}}}}, e^{j\bm{\Phi^{\text{n}}}}, je^{j\bm{\Phi^{\text{n}}}}, -e^{j\bm{\Phi^{\text{n}}}}, -je^{j\bm{\Phi^{\text{n}}}} \right\}$ is the reflection coefficient matrix of the $k$-th time slot. At this point, the interleaving sequence over all $K$ time slots is defined as $\bm{\Gamma}^{\text{i}} = \left(\bm{\Gamma}_{1}^{\text{i}}, \bm{\Gamma}_{2}^{\text{i}}, ..., \bm{\Gamma}_{K}^{\text{i}}  \right)$. Then, the time-domain sequence of electric field strength, $E^{\text{i}} = \left(E_{1}^{\text{i}}, E_{2}^{\text{i}}, ..., E_{K}^{\text{i}}  \right)$, at any target point in space can be derived from Eq.~(\ref{equ_通用场分析}).

The objective of interleaving sequence is to control the spatial distribution of the SNR, i.e., to ensure a high SNR at Bob's location while maintaining low SNR at Eves' locations. The interleaving sequence in the time domain fundamentally operates by injecting artificial noise to regulate the SNR. To intuitively display the level of artificial noise, we adopt the Error Vector Magnitude (EVM) as the metric for assessing secrecy performance. The equivalence between EVM and SNR can be expressed as $\text{EVM} \approx 1/\sqrt{\text{SNR}}$. The EVM characterizes the relative error between the actually received signal and the ideal reference signal, serving as an intuitive indicator of the level of noise. Its formal definition is
%
%
%
%
%
%
%
%
\begin{align}
    \text{EVM}(\bm{\Gamma}) = \sqrt{\frac{\sum_{k=1}^{K} \left| E_{k}-E^{\text{ref}} \right|^{2}}{K \left| E^{\text{ref}} \right| ^{2}}} \times 100\% ,
\end{align}%
where $\bm{\Gamma}$ is the sequence of reflection coefficient matrix, $E_{k}$ is the electric field strength corresponding to $\bm{\Gamma}_{k}$, and $E^{\text{ref}}$ is the ideal reference electric field strength generated by focusing phase matrix $\bm{\Phi}^{\text{f}}$. Therefore, the design objective of EVM is 
%
%
%
%
%
%
%
%
\begin{align}
    & \max_{\bm{\Gamma}} \; \overline{\text{EVM}^{\text{eve}}} \notag \\
    & ~\text{s.t.~} 
    \, \overline{\text{EVM}^{\text{bob}}} < \text{EVM}_{0},
    \label{equ_17_EVM优化建模}
\end{align}
%
%
%
where $\text{EVM}_{0}$ denotes the EVM threshold for the corresponding modulation scheme, as defined in the IEEE 802.11 standard~\cite{STDGT24607}. $\overline{\text{EVM}^{\text{eve}}}$ and $\overline{\text{EVM}^{\text{bob}}}$ represent the average EVM at the sampling points for Eve and Bob, respectively. Specifically, the sampling points for Eve are located in the high gain regions and outer regions in Fig.~\ref{fig_07_零陷模型}, while those for Bob are situated in the nulling region.

Directly optimizing the sequence of reflection coefficient matrix $\bm{\Gamma}$ in Eq.~(\ref{equ_17_EVM优化建模}) poses a significant challenge due to its high complexity. For instance, according to Eq.~(\ref{equ_14_联合时空编码}), if the time slot width is set to $\tau = 2 \mu \mathrm{s}$, the sequence length within $1 \mathrm{s}$ reaches $K=5 \times 10^{5}$. This implies that the problem dimension for a mere $1 \mathrm{s}$ sequence is as high as $5 \times 10^{5}$, making direct optimization extremely difficult. To address this issue and achieve a balance among low computational complexity, high real-time capability, and strong secrecy performance, we propose a time-domain phase interleaving scheme.

\begin{figure}[!t]
    \centering
    \subfloat[Interleaving sequence in time domain]{\includegraphics[width=248pt]{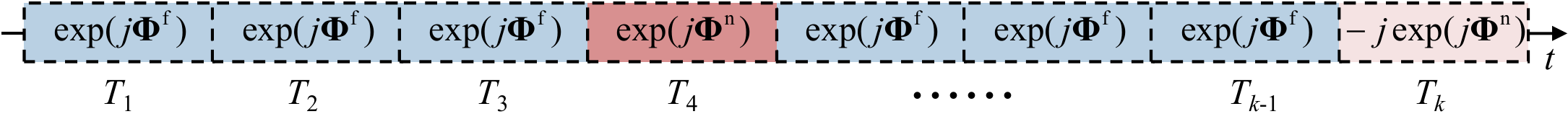}} 
    \label{fig_10a_stc一维时序} \\
    \subfloat[Bob]{\includegraphics[height=71pt]{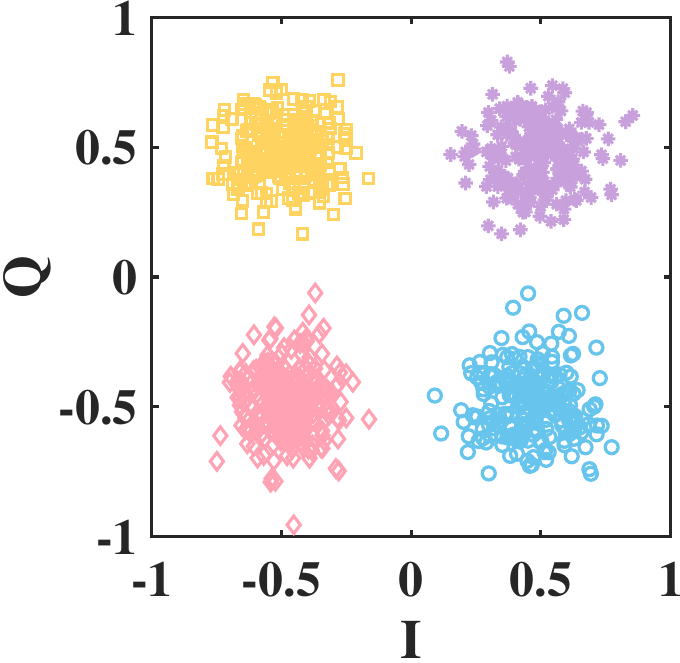}}
    \label{fig_10b_bob}
    \hfil 
    \subfloat[Eve]{\includegraphics[height=71pt]{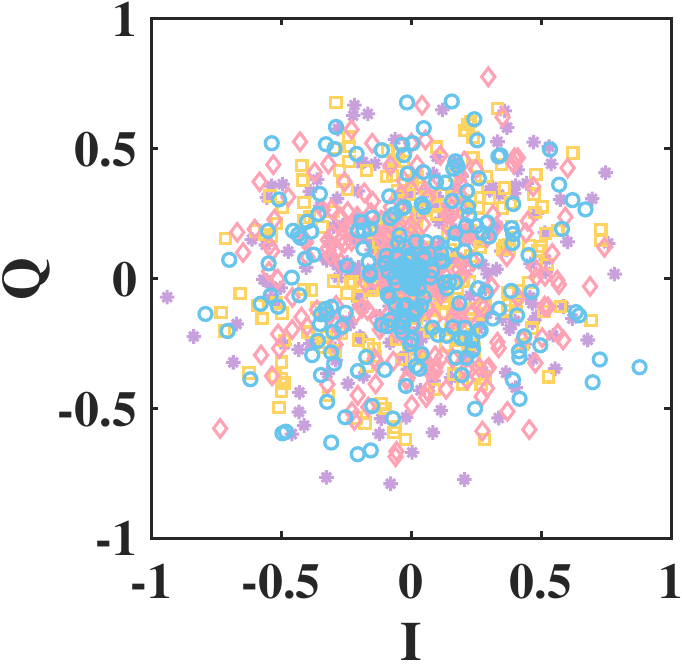}}
    \label{fig_10c_eve} 
    \hfil
    \subfloat[EVM]{\includegraphics[height=71pt]{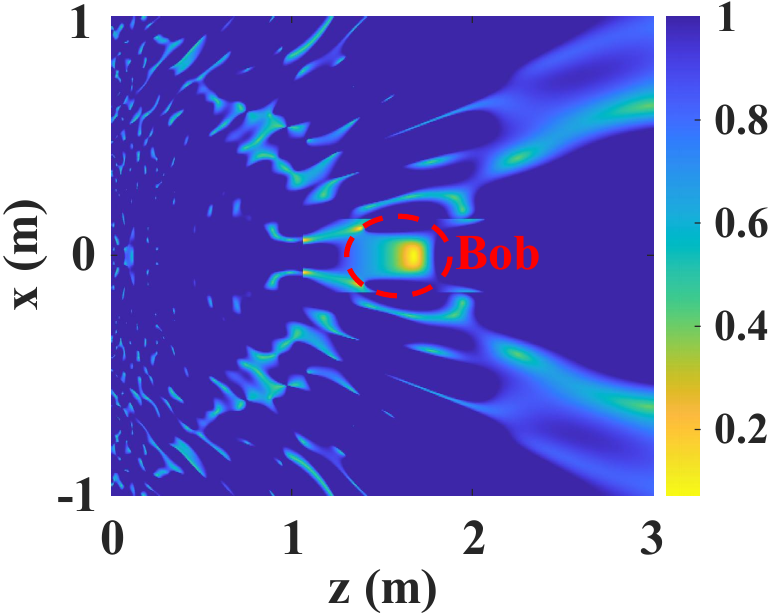}}
    \label{fig_10d_EVM}    
    \caption{The influence of interleaving sequence. (a) 
    Interleaving sequence in time domain. (b) Constellation at Bob. (c) Constellation at Eve. (d) EVM distribution in xoz plane. }
    \label{fig_10总_联合时空编码}
    \vspace{-1.8em}
\end{figure}

The core concept of the time-domain phase interleaving scheme is to first design finite-length perturbed phase sequences that effectively interferes with Eves higher than Bob. These are then expanded by evenly inserting constant phase sequences, forming interleaving sequences until the condition $\overline{\text{EVM}^{\text{bob}}}<\text{EVM}_{0} <\overline{\text{EVM}^{\text{eve}}} $ is met. In practice, a diverse library of pre-generated basic perturbed phase sequences is stored in the RIS controller. This enables the RIS to synthesize interleaving sequences in real-time, dynamically switch basic sequences, and adjust the time-slot ratio between perturbed and constant phase sequences, as shown in Fig.~\ref{fig_10总_联合时空编码}a. This approach not only jams the Eves in real-time but also prevents them from learning and cracking the fixed interference pattern. There are two key challenges in the interleaving sequence design: (i) establishing a validation criterion for basic perturbed phase sequences, and (ii) determining the effective time-slot ratio allocation between perturbed and constant phase sequences. We derive closed-form solutions for these, thus avoiding the prohibitive complexity of direct brute-force optimization.

Consider a perturbed phase sequence of length $K^{\text{null}}$, denoted as $\bm{\Gamma}^{\text{null}} = \left(\bm{\Gamma}_{1}^{\text{null}}, \bm{\Gamma}_{2}^{\text{null}}, ..., \bm{\Gamma}_{K^{\text{null}}}^{\text{null}}  \right)$, where
$\bm{\Gamma}_{k^{\text{null}}}^{\text{null}} \in \left\{ e^{j\bm{\Phi^{\text{n}}}}, je^{j\bm{\Phi^{\text{n}}}}, -e^{j\bm{\Phi^{\text{n}}}}, -je^{j\bm{\Phi^{\text{n}}}} \right\}$. The electric field strength sequence generated by this sequence via Eq.~(\ref{equ_通用场分析}) is $E^{\text{null}} = \left( E_{1}^{\text{null}},E_{2}^{\text{null}},...,E_{K^{\text{null}}}^{\text{null}} \right)$. Consequently, the EVM induced by the basic perturbed phase sequence at the target location $\text{p}$ in space can be expressed as
%
%
%
%
%
%
%
%
\begin{align}
    & \text{EVM}^{\text{null, p}}(\bm{\Gamma}^{\text{null}})
    = \sqrt{\frac{\sum_{k^{\text{null}}=1}^{K^{\text{null}}} \left| E_{k^{\text{null}}}^{\text{null,p}}-E^{\text{ref, p}} \right| ^{2}}{K^{\text{null}} \left| E^{\text{ref, p}} \right|^{2}}}, \notag \\
    & \text{p} \in \left\{ \text{bob, eve} \right\},
    \label{equ_18_噪声基序列EVM}
\end{align}%
where $E^{\text{ref, p}}$ represents the ideal reference electric strength at the point p under the focusing phase matrix $\bm{\Phi}^{\text{f}}$.

Similarly, let a constant phase sequence of length $K^{\text{focus}}$ be defined as $\bm{\Gamma}^{\text{focus}} = \left(\bm{\Gamma}_{1}^{\text{focus}}, \bm{\Gamma}_{2}^{\text{focus}}, ..., \bm{\Gamma}_{K^{\text{focus}}}^{\text{focus}}  \right)$, where $\bm{\Gamma}_{k^{\text{focus}}}^{\text{focus}} \in \left\{ e^{j\bm{\Phi^{\text{f}}}} \right\}$. The time-slot ratio $\frac{K^{\text{focus}}}{K^{\text{null}}}$ is a positive integer. The corresponding electric field strength sequence generated by this constant phase sequence is $E^{\text{focus}} = \left( E_{1}^{\text{focus}},E_{2}^{\text{focus}},...,E_{K^{\text{focus}}}^{\text{focus}} \right)$. As illustrated in Fig.~\ref{fig_10总_联合时空编码}a, the interleaving sequence $\bm{\Gamma}^{\text{inter}}$ is formed by uniformly inserting the constant sequence $\bm{\Gamma}^{\text{focus}}$ into the perturbed sequence $\bm{\Gamma}^{\text{null}}$.
The electric field strength sequence $E^{\text{inter}}$ corresponding to $\bm{\Gamma}^{\text{inter}}$ is denoted as
%
%
%
%
%
%
%
%
\begin{align}
    E^{\text{inter}}= \;( \;
    & E_{1}^{\text{focus}}, E_{2}^{\text{focus}}, ..., E_{\frac{K^{\text{focus}}}{K^{\text{null}}}}^{\text{focus}}, E_{1}^{\text{null}}, \notag\\
    & E_{\frac{K^{\text{focus}}}{K^{\text{null}}}+1}^{\text{focus}}, ..., E_{\frac{2K^{\text{focus}}}{K^{\text{null}}}}^{\text{focus}}, E_{2}^{\text{null}}, \notag\\
    & E_{\frac{2K^{\text{focus}}}{K^{\text{null}}}+1}^{\text{focus}}, ..., E_{K^{\text{focus}}}^{\text{focus}}, E_{K^{\text{null}}}^{\text{null}}
    \;) ,
\end{align}%
where the total length is $K^{\text{inter}}=K^{\text{focus}}+K^{\text{null}}$.
The EVM generated by the interleaving sequence can be calculated as 
%
%
%
%
%
%
%
%
%
%
%
%
\begin{align}
    & \text{EVM}^{\text{inter, p}}(\bm{\Gamma}^{\text{inter}})
    = \sqrt{\frac{\sum_{k^{\text{inter}}=1}^{K^{\text{inter}}} \left| E_{k^{\text{inter}}}^{\text{inter,p}}-E^{\text{ref, p}} \right| ^{2}}{K^{\text{inter}} \left| E^{\text{ref, p}} \right|^{2}}} \notag \\
   & = \sqrt{\frac{ \sum_{k^{\text{focus}}=1}^{K^{\text{focus}}} \left| E_{k^{\text{focus}}}^{\text{focus,p}}-E^{\text{ref, p}} \right| ^{2}+\sum_{k^{\text{null}}=1}^{K^{\text{null}}} \left| E_{k^{\text{null}}}^{\text{null,p}}-E^{\text{ref, p}} \right| ^{2}}{K^{\text{inter}} \left| E^{\text{ref, p}} \right|^{2}}}, \notag \\
    & \text{p} \in \left\{ \text{bob, eve} \right\}.
    \label{equ_21_联合编码EVM_原始}
\end{align}%
Since $E_{k^{\text{focus}}}^{\text{focus,p}}$ and $E^{\text{ref, p}}$ are both generated by the focusing phase matrix $\bm{\Phi}^{\text{f}}$ at location $\text{p}$, it follows that $E_{k^{\text{focus}}}^{\text{focus,p}}-E^{\text{ref, p}}=0$. With reference to Eq.~(\ref{equ_18_噪声基序列EVM}), the Eq.~(\ref{equ_21_联合编码EVM_原始}) can be simplified to 
%
%
%
%
%
%
%
%
%
%
%
%
\begin{align}
    & \text{EVM}^{\text{inter, p}}(\bm{\Gamma}^{\text{inter}})
    = \sqrt{\frac{K^{\text{null}}}{K^{\text{focus}}+K^{\text{null}}}}\text{EVM}^{\text{null, p}} \notag \\
   & \text{p} \in \left\{ \text{bob, eve} \right\}.
    \label{equ_22_联合编码EVM_简化}
\end{align}
%
%
The Eq.~(\ref{equ_22_联合编码EVM_简化}) indicates the relationship between $\text{EVM}^{\text{inter, p}}$ and $\text{EVM}^{\text{null, p}}$. This relationship is crucial for subsequently deriving the closed-form validation criterion for basic perturbed phase sequences, and establishing the upper and lower bounds for the time-slot ratio $\frac{K^{\text{focus}}}{K^{\text{null}}}$ allocation between perturbed and constant phase sequences in the interleaving sequence.

Based on the requirements outlined in Eq.~(\ref{equ_17_EVM优化建模}), the EVM of interleaving sequence for each location must satisfy the following constraint
%
%
%
%
%
%
%
%
\begin{align}
\begin{cases}
    \overline{\text{EVM}^{\text{inter, bob}}} = \sqrt{\frac{K^{\text{null}}}{K^{\text{focus}}+K^{\text{null}}}}\overline{\text{EVM}^{\text{null, bob}}} < \text{EVM}_{0} \\
    \overline{\text{EVM}^{\text{inter, eve}}} = \sqrt{\frac{K^{\text{null}}}{K^{\text{focus}}+K^{\text{null}}}}\overline{\text{EVM}^{\text{null, eve}}} > \text{EVM}_{0}
\end{cases}.
\label{equ_23_联合编码EVM具体约束}
\end{align}%
Rearranging Eq.~(\ref{equ_23_联合编码EVM具体约束}) yields the effective range for the time-slot ratio as
%
%
%
%
%
%
%
%
\begin{align}
\begin{cases}
    \frac{K^{\text{focus}}}{K^{\text{null}}} \in \left( 
    \left( \frac{\overline{\text{EVM}^{\text{null, bob}}}}{\text{EVM}_{0}} \right)^{2}-1, 
    \left( \frac{\overline{\text{EVM}^{\text{null, eve}}}}{\text{EVM}_{0}} \right)^{2}-1
    \right) \\
    \frac{K^{\text{focus}}}{K^{\text{null}}} \in \mathbb{Z}^{+}
\end{cases},
    \label{equ_24_联合编码信号噪声时间占比}
\end{align}%
where $\mathbb{Z}^{+}$ denotes the set of positive integers. The non-negativity constraint $\frac{K^{\text{focus}}}{K^{\text{null}}} \geq 0$ requires the upper bound of the interval in Eq.~(\ref{equ_24_联合编码信号噪声时间占比}) to be positive, so we can get the condition $\overline{\text{EVM}^{\text{null, eve}}} > \text{EVM}_{0}$. This indicates that the average EVM of the perturbed phase sequence at eves' locations must exceed the EVM threshold of the corresponding modulation scheme. Furthermore, for the interval in Eq.~(\ref{equ_24_联合编码信号噪声时间占比}) to be valid, the upper bound must be greater than the lower bound, therefore we can get an another condition $\overline{\text{EVM}^{\text{null, eve}}} > \overline{\text{EVM}^{\text{null, bob}}} $. This condition shows that the average EVM of perturbed phase sequence at Eves' locations should be greater than at bob's, and demonstrates the necessity of beam nulling pattern. Synthesizing these two necessary conditions, the closed-form validation criterion for a basic perturbed phase sequence is given by
%
%
%
%
%
%
%
%
\begin{align}
    \overline{\text{EVM}^{\text{null, eve}}} >
    \max \left\{\overline{\text{EVM}^{\text{null, bob}}}, \;
    \text{EVM}_{0}
    \right\}.
    \label{equ_25_有效噪声基序列判定标准}
\end{align}%
Any basic perturbed phase sequence that satisfies the EVM constraint specified in Eq.~(\ref{equ_25_有效噪声基序列判定标准}) is deemed effective.

\begin{algorithm}[!t]
    \caption{Secure Location Modulation of NF-SecRIS}
    \label{algo_02_系统总体流程}
    \begin{algorithmic}[1]
        \REQUIRE
        $\text{EVM}_{0}$ and the entire input of Algorithm \ref{algo_01_null}
        \ENSURE
        Interleaving Sequence $\bm{\Gamma}^{\text{inter}}$
        \STATE \textbf{Beam Focusing Module:}
        \STATE Calculate $\bm{\Phi}^{\text{f}}$ through near-field phase compensation via (\ref{eq_02_focus}) and (\ref{equ_量化规则});
        \STATE Generate the basic constant phase sequence $\bm{\Gamma}^{\text{focus}}$;
        \STATE \textbf{Beam Nulling Module:}
        \STATE Calculate $\bm{\Phi}^{\text{n}}$ through Squeeze-Nulling Model via Algorithm \ref{algo_01_null};
        \STATE Generate the basic perturbed phase sequence $\bm{\Gamma}^{\text{null}}$ satisfying (\ref{equ_25_有效噪声基序列判定标准}); 
        \STATE \textbf{Interleaving Module:} 
        \STATE Calculate the time-slot ratio $\frac{K^{\text{focus}}}{K^{\text{null}}}$ via (\ref{equ_24_联合编码信号噪声时间占比});
        \STATE Generate the dynamic interleaving sequence $\bm{\Gamma}^{\text{inter}}$ by randomly switching the $\bm{\Gamma}^{\text{null}}$ and $\frac{K^{\text{focus}}}{K^{\text{null}}}$ in real time; 
        \RETURN $\bm{\Gamma}^{\text{inter}}$.
    \end{algorithmic}
\end{algorithm}%


In summary, a key feature of the proposed SLM-based NF-SecRIS system is its inherent independence from the transmitter and receiver, operating without any synchronization requirements with them. Furthermore, it works without relying on extra communication or signal processing protocols. This self-sufficient approach ensures that the 2D PLS system achieves low complexity and strong robustness. The overall workflow is summarized in Algorithm~\ref{algo_02_系统总体流程}.

\vspace{-1.2em}
\section{Implementation}

\begin{figure}[!t]
    \centering
    \includegraphics[width=250pt]{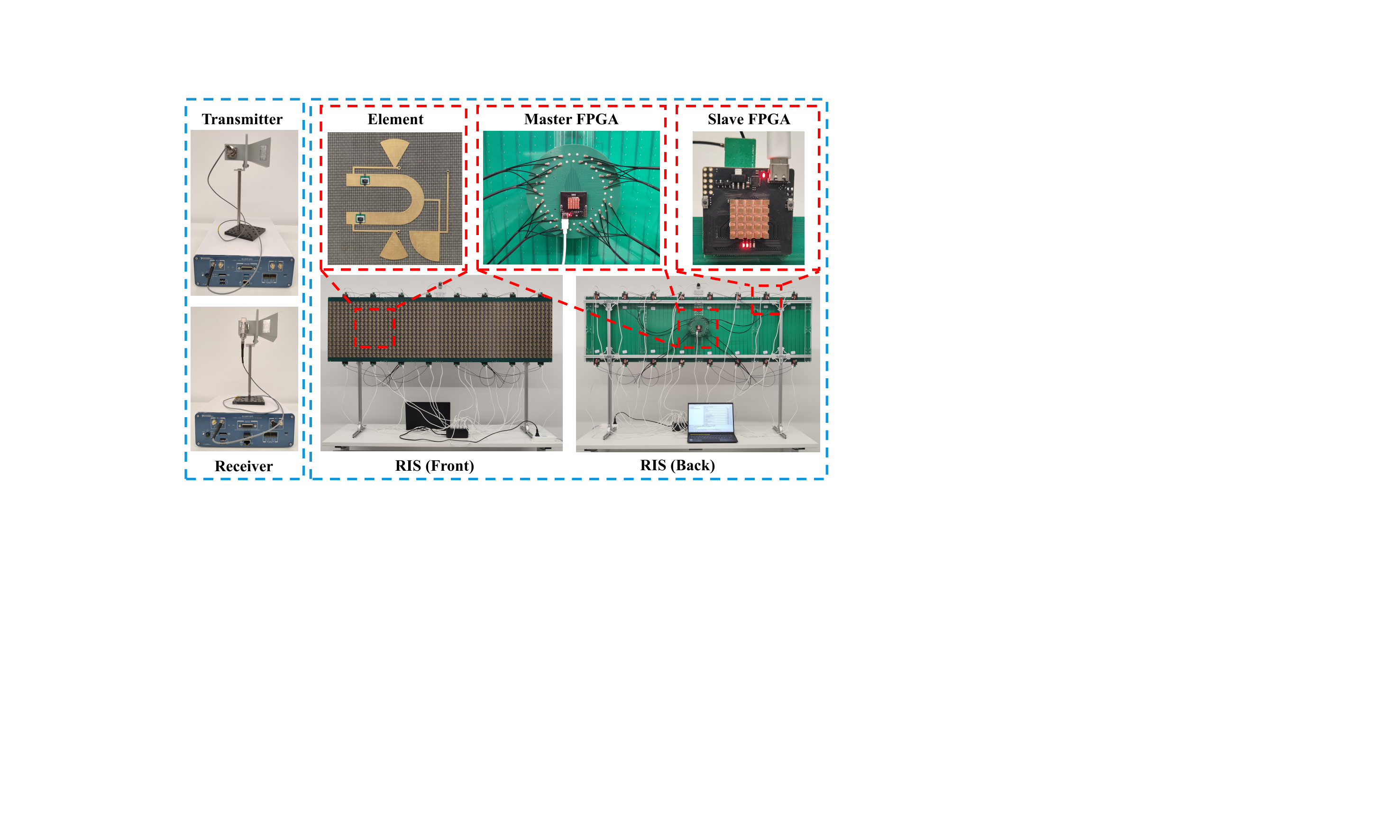}
    \caption{Prototype of NF-SecRIS.}
    \label{fig_11_实现图}
    \vspace{-1.8em}
\end{figure}

In this section, we present the system implementation of NF-SecRIS, as illustrated in Fig.~\ref{fig_11_实现图}.

\vspace{-1.2em}
\subsection{RIS Array}

We design a 2-bit ultra-large-scale RIS array operating at 5.8 GHz with $14 \times 56$ unit cells. Each unit cell has a side length of 27.8 mm, resulting in a total aperture size of $39~\text{cm} \times 156~\text{cm} $. The unit cell structure is adapted from the design presented in~\cite{9522074}, with appropriate parameter optimization. Each unit cell incorporates two PIN diodes with switching voltages of 1 V (ON) and 0 V (OFF), and the model is Skyworks-SMP1340-040LF. For ease of fabrication, the entire array is subdivided into four subarrays with $14 \times 14$ elements. Each subarray is manufactured individually and subsequently integrated together.

\vspace{-1.2em}
\subsection{FPGA Control}

%
%
%

For RIS control, we develop a master-slave FPGA architecture to independently and synchronously control each RIS unit. The FPGA chip model used is Altera-EP4CE6F17C8N. The precomputed spatial phase matrices and time-domian basic sequences are pre-stored in the slave FPGAs to serve as the foundational states for real-time random invocation. The overall workflow is as follows: the host computer generates real-time state-switching commands and transmits them to the master FPGA, which then distributes the instructions to the slave FPGAs via a command bus, while a clock signal ensures synchronous operation across all slave FPGAs. Each slave FPGA outputs high/low logic levels according to the instructions, controlling the states of the RIS diodes with microsecond-scale switching rates to achieve dynamic and synchronized spatial-temporal phase patterns generation in real-time.

%
%
%
%
\begin{figure*}[!t]
    \centering
    \subfloat[Deployment layout]{\includegraphics[height=175pt]{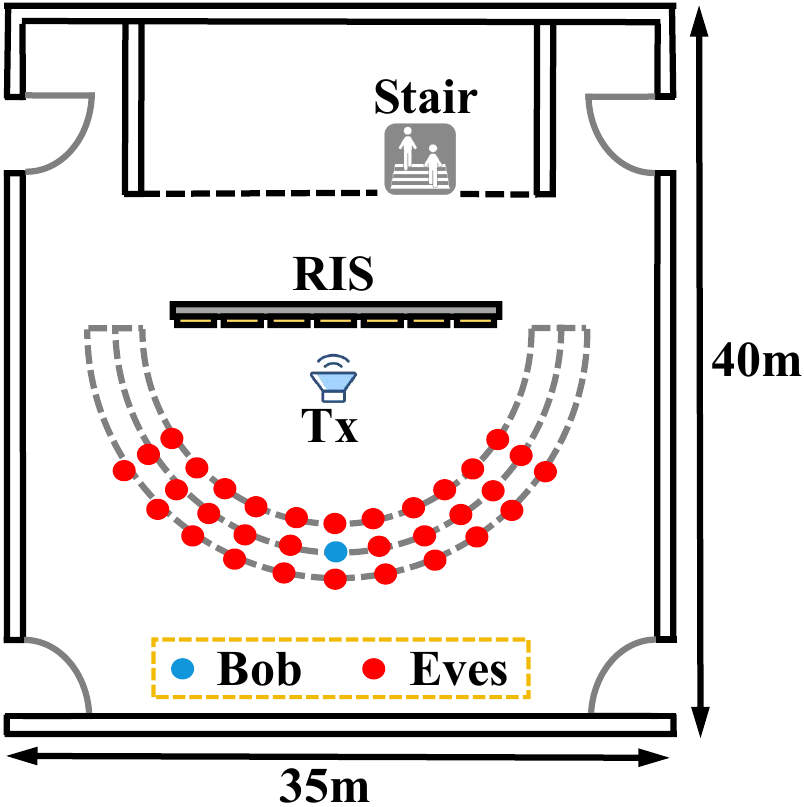}}
    \label{fig_12a_实验布局图}
    \hfil
    \subfloat[Evaluation Scenario]{\includegraphics[height=175pt]{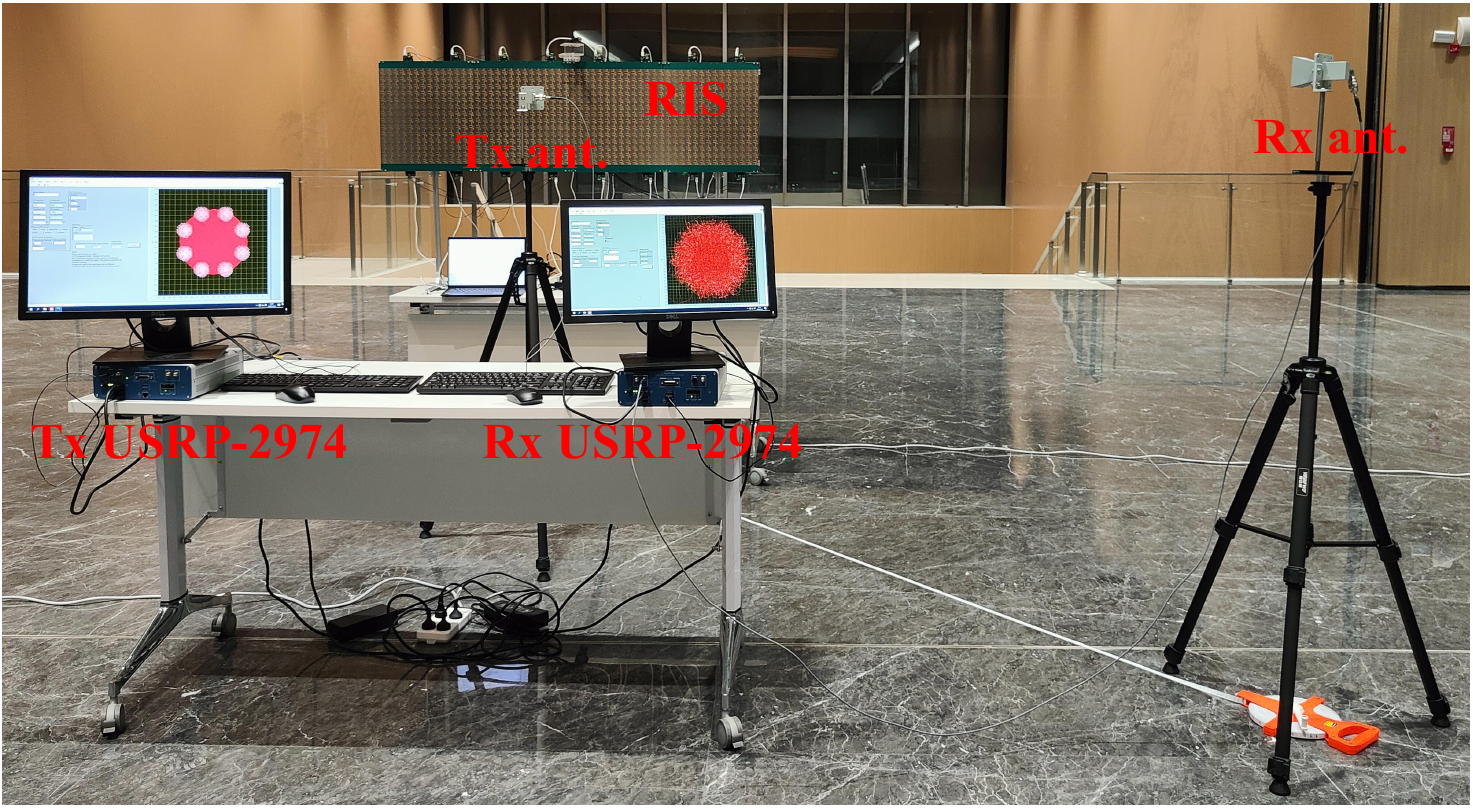}}
    \label{fig_12b_场景实拍图}  
    \caption{Experiment Setup.}
    \label{fig_12总_实验Setup}
    \vspace{-1.8em}
\end{figure*}%

Each slave FPGA controls a region of $7 \times 7$ unit cells. Each subarray board employs four slave FPGAs, resulting in a total of 16 slave FPGAs for the entire array. To ensure scalability of the slave FPGA count, we design a master-slave adapter board. The master FPGA feeds both command and clock signals into this adapter board, which then distributes these signals to each slave FPGA. Since the number of signal interfaces on the adapter board can be expanded almost without limit, the number of slave FPGAs, and therefore the scale of the RIS array, can be extended almost without limit. This design facilitates the construction of ultra-large-scale RIS array, thereby supporting extensive near-field communication coverage.

\vspace{-1em}
\subsection{Transmitter and Receiver}

We employ two USRP-2974 devices as the transmitter and receiver, respectively, each equipped with a horn antenna. The transmitter emits modulated signals such as ASK, FSK, PSK, and QAM through the TX1 channel. After being reflected by the RIS, the signals are captured by the receiving antenna and fed into the RX2 channel of the receiver. Finally, through signal processing, key performance metrics including the received constellation diagram and bit error rate are obtained.

\section{Evaluation}

In this section, we experimentally validate the 2D PLS performance of NF-SecRIS. We assume that NF-SecRIS does not know the numbers or locations of eavesdroppers. Both legitimate users and eavesdroppers possess identical prior knowledge and capabilities.

\vspace{-1.2em}
\subsection{Experiment Setup}

We conduct the secure communications experiments in a hall measuring $\text{40m}\times\text{35m}$, as shown in Fig.~\ref{fig_12总_实验Setup}. Specifically, Fig.~\ref{fig_12总_实验Setup}a illustrates the layout diagram, while Fig.~\ref{fig_12总_实验Setup}b shows the scenario of the actual setup. Here, we continue to use the coordinate system introduced in Fig.~\ref{fig_04_focus_model}, where the center of the RIS is set as the coordinate origin, the RIS lies in the $xoy$ plane, the $z$ axis is normal to the array surface, and the $xoz$ plane represents the horizontal plane. By default, we focus on the two-dimensional security performance within the $xoz$ plane. For convenience, a supplementary polar coordinate system is adopted to specify the positions of spatial sampling points within the $xoz$ plane. For example, in Fig.~\ref{fig_12总_实验Setup}a, the blue dot indicates Bob's location with polar coordinates $(1.6\text{m},0^{\circ})$, and Tx indicates the feed horn's position at polar coordinates $(0.8\text{m},0^{\circ})$. In subsequent experiments, spatial sampling points are represented using this polar coordinate system in the $xoz$ plane by default. Moreover, the spatial sampling region spans distances from $1\text{m}$ to $3\text{m}$ and angles from $-90^{\circ}$ to $90^{\circ}$, with initial sampling intervals of $0.2\text{m}$ and $10^{\circ}$, respectively. In the vicinity of Bob, the sampling grid is denser using refined intervals of $0.1\text{m}$ in distance and $5^{\circ}$ in angle to capture detailed performance variations.

In the communication links, the key parameters are configured as follows. At the transmitter, the carrier frequency is 5.8 GHz, the modulation scheme is 8PSK, and the symbol rate is 125 kHz, resulting in a symbol duration of $8\mu \text{s}$. At the RIS, the switching rate of the diodes is set to 500 kHz, corresponding to a pulse slot width of $2\mu \text{s}$. The ratio of constant phase time slots to perturbed phase time slots in the interleaving phase sequence is $\frac{K^{\text{focus}}}{K^{\text{null}}}=3$. At the receiver, spatial sampling across the $xoz$ plane is performed by moving the receiving antenna, with its main lobe consistently aligned toward the center of the RIS. Unless otherwise specified, the aforementioned configurations are adopted by default in the subsequent experiments.

\vspace{-1.2em}
\subsection{Security Metric}

To validate the 2D secure communication capability of NF-SecRIS, we employ the bit error rate (BER) as the unified security metric. It is important to note that a BER of $50\%$ indicates complete demodulation failure at the receiver, implying that it can only randomly guess the received symbols. For example, in QPSK, each symbol carries 2 bits, with four possible symbol values: $\{00,01,10,11\}$. Suppose the transmitted symbol is $00$, the corresponding number of error bits for the four possible guessed symbols would be $\{0,1,1,2\}$. If the receiver guesses randomly with equal probability (0.25 for each case), the weighted average number of bit errors is $0.25\times(0+1+1+2)=1$, resulting in a BER of $1 \div 2 = 50\%$. Specifically, we use the cumulative BER after transmitting 100 Mb of data. In subsequent experiments, we will demonstrate that NF-SecRIS can force the BER above $40\%$ for most modulation schemes at eavesdropper positions, leading to complete demodulation failure, while legitimate user maintains normal demodulation performance.

\vspace{-1.2em}
\subsection{2D Secure Communication}

\begin{figure}[!t]
    \centering
    \includegraphics[width=200pt]{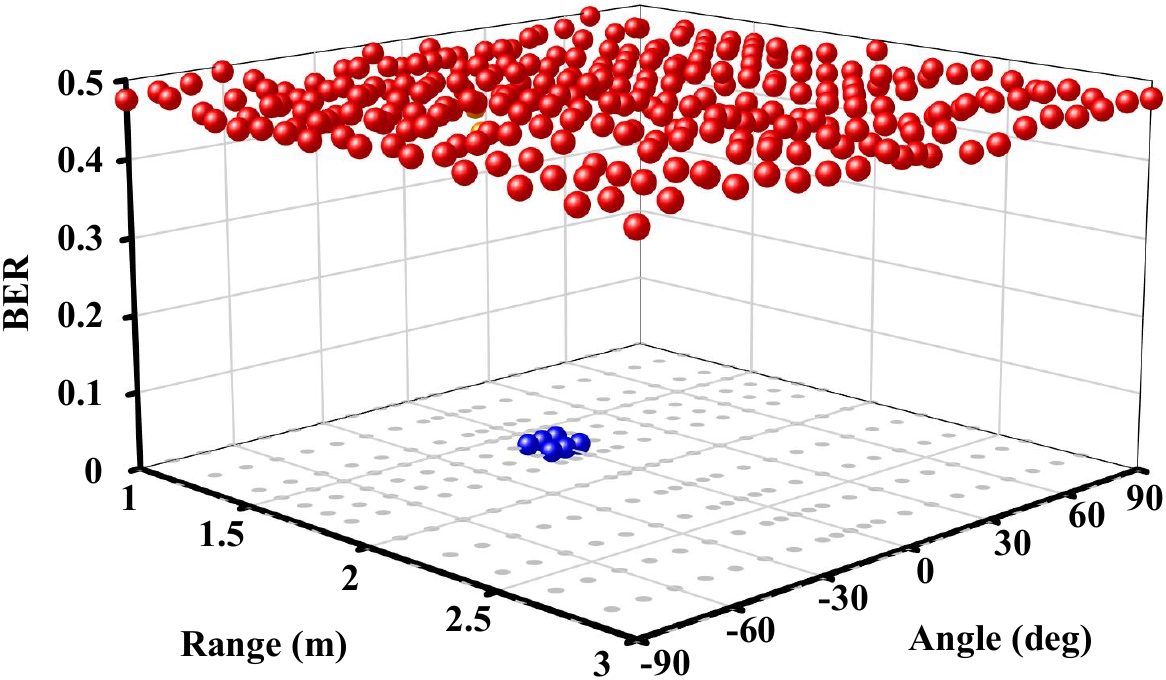}    
    \caption{2D PLS performance of NF-SecRIS.}
    \label{fig_13总_测试_二维保密}
    \vspace{-1.8em}
\end{figure}

To verify that NF-SecRIS can achieve 2D PLS near-field communications, we sample the BER distribution in the $xoz$ plane under a default experiment setup. Fig.~\ref{fig_13总_测试_二维保密} presents the measurement results of BER distribution. It shows that the target location for Bob exhibits a BER of $7.83 \times 10^{-5}$. The surrounding low-BER region is defined as the secure communication area, which is measured to be within $r \in [1.6, 1.7]\text{m} \times \theta \in [-5, 5]^{\circ}$, with the BER consistently remaining below the $10^{-4}$ level. In contrast, Eves located outside this area experience BERs exceeding $40\%$, indicating complete demodulation failures and unsuccessful communications. In summary, the results confirm that NF-SecRIS successfully achieves 2D PLS in near-field communications.

\vspace{-1.2em}
\subsection{Noise Is Necessary}

%
%
%

\begin{figure}[!t]
    \centering
    \includegraphics[width=250pt]{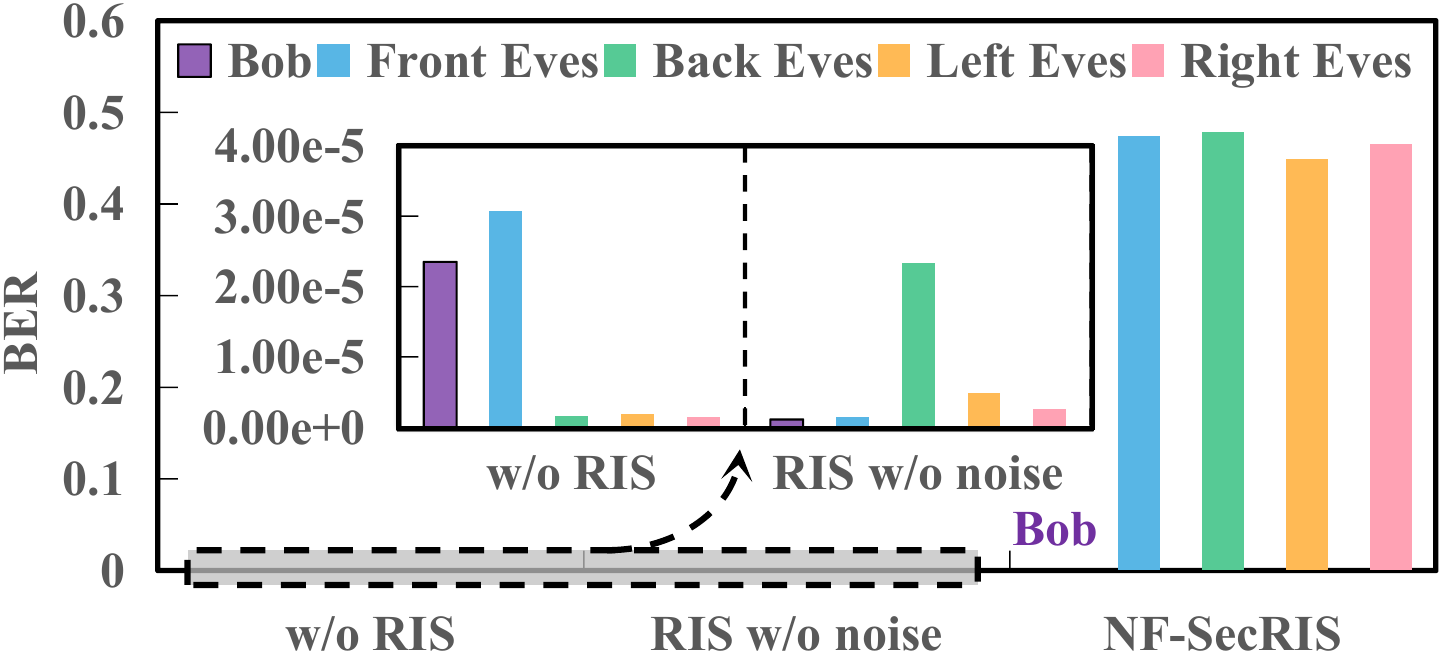}
    \caption{Noise is necessary for 2D PLS.}
    \label{fig_14_技术必要性}
    \vspace{-1.2em}
\end{figure}

The artificial noise generated by perturbed phase sequence serves as the core component of NF-SecRIS for achieving secure communication. To validate its necessity, an ablation experiment is designed. The experiment employs the default sampling setup and communication link configuration. Fig.~\ref{fig_14_技术必要性} presents the test results of average BER within each zone which is aforementioned in Fig.~\ref{fig_07_零陷模型}. The results indicate that when the RIS is inactive, both Bob and Eves can maintain low BERs, meaning Eves can successfully eavesdrop the communication data. When the RIS performs only constant phase sequence, without artificial noise generated by perturbed phase sequence, the BERs of Eves remain extremely low, allowing successful eavesdropping in 2D region. The reason is that the constellation at Eve remains undistorted, and the Eve with sufficient sensitivity can still demodulate the signal. In contrast, the results of NF-SecRIS demonstrate that only Bob can achieve stable communication, while eavesdroppers at other locations experience BER exceeding $40\%$, leading to eavesdropping failure. Consequently, these experimental results confirm that the artificial noise generated by perturbed phase sequence is an indispensable module in NF-SecRIS. Its function is to disrupt the constellations of eavesdroppers, causing demodulation failure and thereby ensuring the performance of 2D secure communications.

\vspace{-1.2em}
\subsection{Security for Different Modulation Schemes}

\begin{figure}[!t]
    \centering
    \includegraphics[width=250pt]{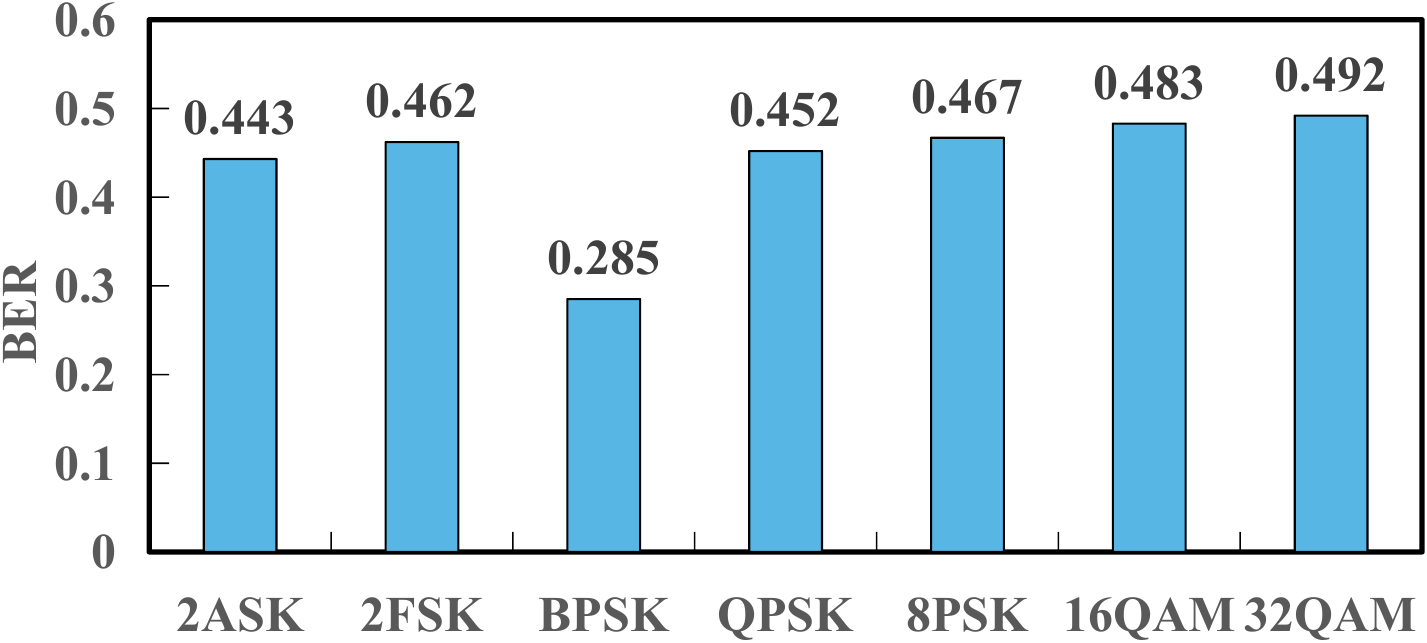}
    \caption{The BER of Eves for different modulation schemes.}
    \label{fig_15_测试_调制方式}
    \vspace{-1.2em}
\end{figure}

To evaluate the security for different modulations, we conduct 2D secure communication experiments using various modulation schemes, including ASK, FSK, PSK, and QAM. All other parameters of communication link retain their default configurations. The sampling points for Bob and Eve remain consistent with previous experiments, and the BER of Eve is averaged for analysis. Fig.~\ref{fig_15_测试_调制方式} presents the measured results of BER for Eve under each modulation scheme. The results demonstrate that NF-SecRIS effectively generates significant interference at eavesdropper locations across all tested modulations. It is noteworthy that BPSK exhibits a relatively lower BER of $28.5\%$, which can be attributed to its lower modulation order and consequent higher tolerance to interference. Nevertheless, this BER level of $28.5\%$ can still prevent Eve from eavesdropping. Furthermore, higher-order modulation schemes show greater sensitivity to the interference, with BER exceeding $40\%$, indicating complete demodulation failure. It is important to highlight that the BER of Bob remains consistently below the $10^{-4}$ level under all modulation schemes. The BER of Bob is not displayed in the figure because of the substantial difference in BER magnitude between Bob and Eve. In conclusion, these findings confirm that NF-SecRIS achieves robust 2D secure communication across a wide range of commonly used modulation techniques.

\vspace{-1.2em}
\subsection{Impact of RIS Time Slot Width}

%
%
%

%
%
%

\begin{figure}[!t]
    \centering
    \includegraphics[width=250pt]{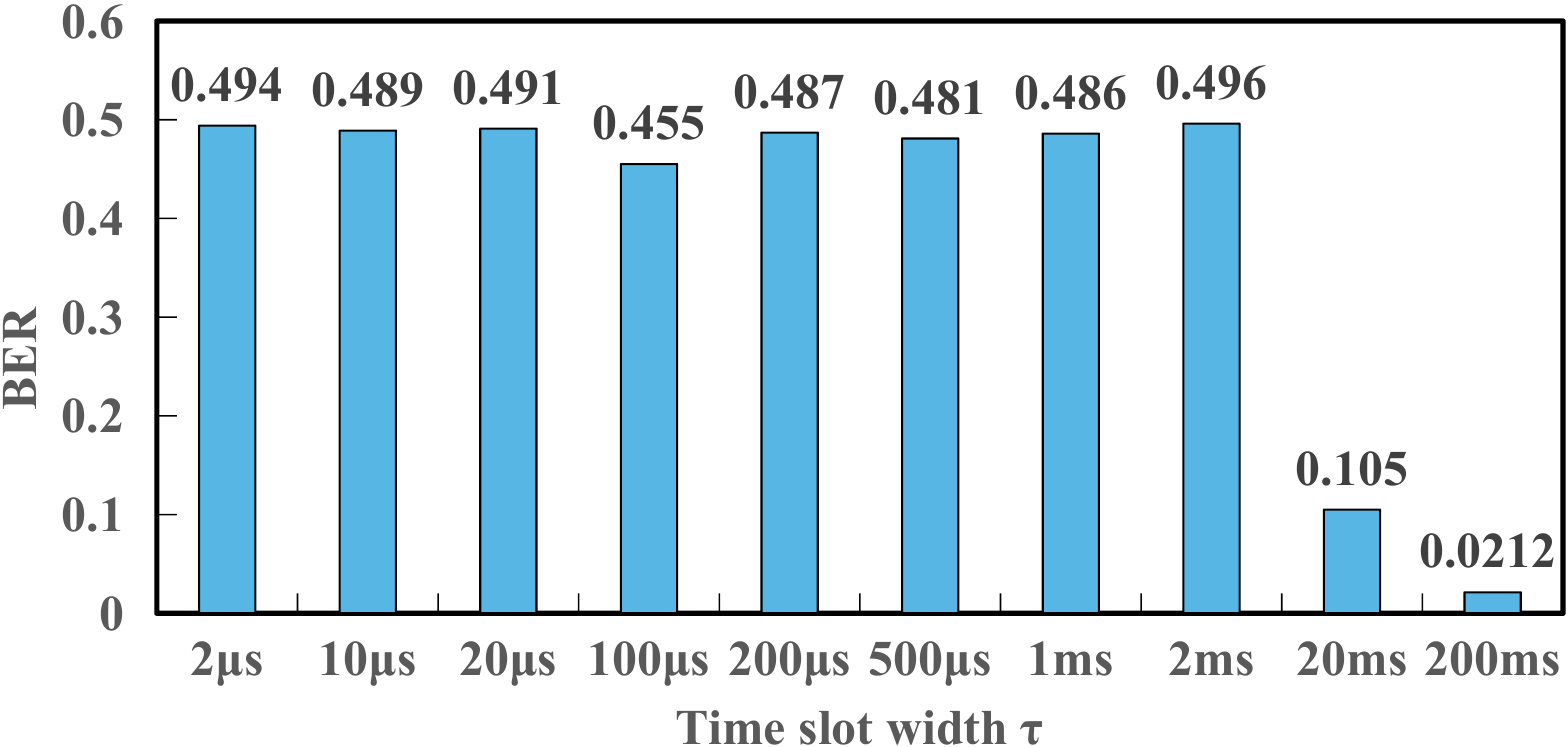}
    \caption{The BER of Eves for different RIS time slot width.}
    \label{fig_16_测试_ris时隙宽度tau}
\end{figure}

A core technology of NF-SecRIS is perturbed phase sequence, for which the RIS time slot width $\tau$ is a critical parameter. This experiment investigates the impact of $\tau$ on the performance of 2D secure communication. The communication link uses the default configuration except for the RIS slot width. It is noteworthy that the symbol rate of the modulated signal is $125\text{kHz}$, corresponding to a symbol duration of $8\mu\text{s}$. The spatial sampling points for Bob and Eve follow the default setting. As shown in Fig.~\ref{fig_16_测试_ris时隙宽度tau}, when $\tau \leq 2\text{ms}$, Eve experiences serve interference. However, when $\tau \geq 20\text{ms}$, the BER of Eve decreases sharply, enabling successful eavesdropping. This is because when the RIS time slot is sufficiently long relative to the symbol duration, the RIS reflection coefficients remain stable within a single symbol, allowing the receiver to track the phase and demodulate the signal. To reliably achieve 2D secure communication, we recommend that the time slot width of NF-SecRIS be comparable to the symbol duration of the modulated signal.

\vspace{-1.2em}
\subsection{Impact of Time-Slot Ratio Allocation}

%
%
%

\begin{figure}[!t]
    \centering
    \includegraphics[width=250pt]{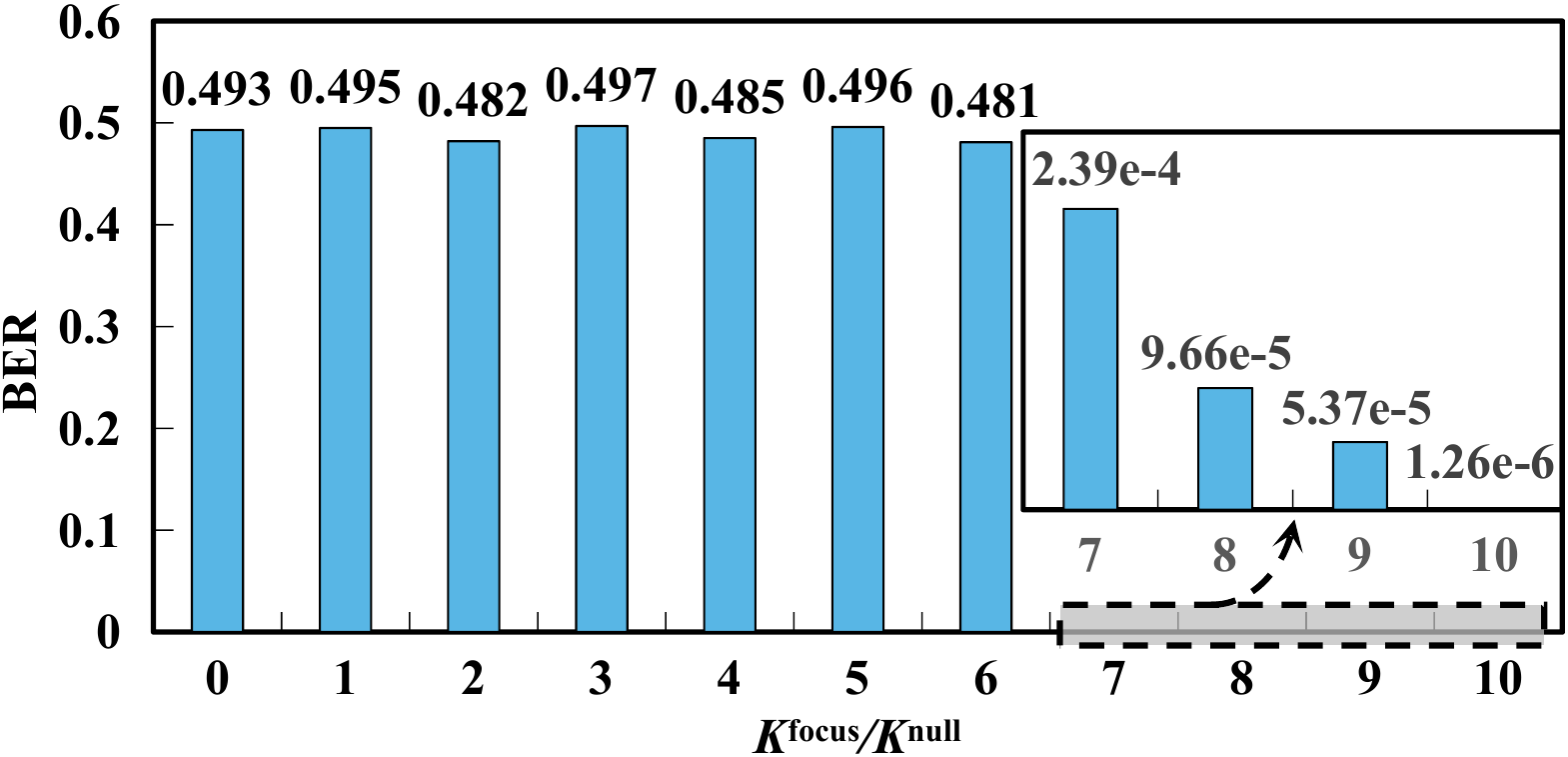}
    \caption{The impact of the ratio of the number of time slots.}
    \label{fig_17_时隙数量比}
    \vspace{-1.8em}
\end{figure}

In the NF-SecRIS framework, another critical parameter in interleaving sequence is the ratio of the number of constant phase time slots to the number of perturbed phase time slots, denoted as $\frac{K^{\text{focus}}}{K^{\text{null}}}$. This experiment validates its specific impact on the interference efficacy against Eve. The communication link employs the default configuration except for the $\frac{K^{\text{focus}}}{K^{\text{null}}}$. The spatial sampling locations remain consistent with previous experiments. Fig.~\ref{fig_17_时隙数量比} presents the BER of Eve under different $\frac{K^{\text{focus}}}{K^{\text{null}}}$. The results demonstrate that when $\frac{K^{\text{focus}}}{K^{\text{null}}} \leq 6$, the BER of Eve remains exceeding $40\%$, indicating complete demodulation failure. Conversely, when $\frac{K^{\text{focus}}}{K^{\text{null}}} \geq 7$, the BER of Eve drops sharply, enabling successful eavesdropping. As $\frac{K^{\text{focus}}}{K^{\text{null}}}$ increases, the EVM monotonically decreases. Once the EVM drops below a specific threshold, the receiver succeeds in demodulation, resulting in a rapid reduction in the BER. Therefore, to effectively jam the eavesdropper, the $\frac{K^{\text{focus}}}{K^{\text{null}}}$ must reside within the valid range specified by Eq.~(\ref{equ_24_联合编码信号噪声时间占比}).

In summary, the previous experiments confirm that NF-SecRIS enables robust 2D PLS near-field communications.

%
%
%

\vspace{-0.5em}
\section{Related Work}



\textbf{Physical Layer Security.} Most experimentally validated PLS systems to date achieve security only in the angular dimension, while lacking protection in the range dimension~\cite{10.1145/3495243.3560547, s41467-025-60725-1, s41467-025-63326-0, 10.1038/s41467-024-50482-y, 10.1038/s41928-021-00664-z, 10.1063/5.0132854,10274844}. RIS-assisted schemes that achieve PLS specifically within the angular domain are proposed in~\cite{10.1145/3495243.3560547},~\cite{s41467-025-60725-1} and ~\cite{10.1038/s41467-024-50482-y}. Secure millimeter-wave communication is realized using space-time-modulated antenna array~\cite{10.1038/s41928-021-00664-z}. Furthermore, secure communication via transmission along multiple distinct paths is explored in \cite{s41467-025-63326-0}, \cite{10.1063/5.0132854} and \cite{10274844}. 

To achieve range-dimensional PLS, RIS-assisted near-field NOMA communication is investigated with an emphasis on evaluating security performance in the range dimension \cite{10902048}. Studies focusing on RIS-assisted near-field OAM and UAV secure communications can be found in \cite{10910063} and \cite{11121378}, respectively. However, these works are predominantly limited to scenarios with a single eavesdropper and remain at the stage of simulation analysis. In contrast, the proposed NF-SecRIS framework addresses the more challenging scenario of multiple eavesdroppers without prior knowledge of their locations, and implements a practical near-field secure communication system.


\textbf{Near-Field Communications.} With the advancement of ultra-large-scale array technology, it is anticipated that a significant portion of daily communication scenarios will operate within the near-field region of arrays~\cite{10541333}. Consequently, near-field communication is poised to become a key technology for 6G~\cite{10716601},~\cite{10380596},~\cite{10934753},~\cite{10944643},~\cite{10.1038/s41467-025-56209-x}. Current studies are predominantly concentrated in two areas: channel modeling and analysis for the near-field region~\cite{11059909},~\cite{10684477},~\cite{10869299},~\cite{10123941}, and the synthesis of advanced beam focusing~\cite{10.5555/3388242.3388317},~\cite{9866003},~\cite{10988518},~\cite{10587118}. However, beam focusing alone is insufficient for unlocking the full potential of range-dimensional manipulation in the near-field. In contrast, NF-SecRIS not only achieves precise beam focusing but also realizes near-field beam nulling at very low computational complexity, thereby simultaneously satisfying the specific requirements for signal and artificial noise distribution.


\textbf{Reconfigurable Intelligent Surface.} With continuous advancements in RIS, it has begun to exhibit multi-functional capabilities~\cite{10833623},~\cite{10720877},~\cite{10718344},~\cite{11008738}. To support the multi-functional capabilities, the RIS controls its electromagnetic properties by manipulating the current distribution via the ON/OFF states of diodes within its FPGA-based control unit~\cite{10159567}. In prior studies, due to constraints on the PIN resource of a single FPGA, RIS typically employs row-column addressing, which precludes independent control of each unit cell~\cite{8917926}. To overcome this limitation and achieve independent, synchronous control over every unit cell in an ultra-large-scale array, we design and implement a master-slave FPGA architecture for NF-SecRIS. Critically, this architecture allows for theoretically unlimited scalability by increasing the number of slave FPGAs, meaning the scale of the array can be expanded virtually without bound. This provides a crucial hardware foundation for advanced near-field communication systems.

\vspace{-0.5em}
\section{Conclusion}
In this paper, we design and fabricate NF-SecRIS, the first ultra-large-scale RIS-based system enabling 2D physical layer secure near-field communications. Moreover, our proposed secure location modulation scheme decouples the spatial-temporal phase patterns design into spatial phase matrix synthesis and temporal sequence design, significantly reducing system complexity. Additionally, we develop a squeeze-nulling model that reduces the dimensionality of near-field null phase matrix optimization from nearly $10^{3}$ to merely 8. Furthermore, we establish a closed-form validation criterion for basic perturbed phase sequences, and construct a basic sequence library which is pre-stored in the RIS controller for real-time deployment. Subsequently, we develop closed-form upper and lower bounds for the time-slot ratio allocation in the interleaving sequences, ensuring real-time generation of spatial-temporal phase patterns. Consequently, our experimental results demonstrate that NF-SecRIS can achieve 2D PLS near-field communications across multiple modulation schemes, such as FSK, ASK, PSK, and QAM. Furthermore, the system can extend its secure near-field communication region by expanding the array aperture. Although our experiments are conducted at 5.8 GHz, the scheme of NF-SecRIS is directly extensible to other frequency bands. In summary, we anticipate that this work will provide valuable insights and serve as a foundational reference for the development of secure communications in 6G wireless systems.



\bibliographystyle{IEEEtran}
\bibliography{02_Section/11_Reference}


\end{document}